\documentclass[prx,amsmath,amssymb,aps,superscriptaddress,twocolumn,longbibliography,nofootinbib]{revtex4-2}

\usepackage{epsfig}
\usepackage{url}
\usepackage{hyperref}
\usepackage{natbib}

\usepackage[normalem]{ulem} 

\usepackage{latexsym}
\usepackage{epsfig}
\usepackage{amsmath}

\usepackage{amssymb}
\usepackage{wasysym}
\usepackage{graphicx}
\usepackage{dcolumn}
\usepackage{verbatim}
\usepackage{enumerate,mdwlist}
\usepackage{bm}
\usepackage[titletoc]{appendix}

\usepackage{amsfonts}
\usepackage{fancyvrb} 
\usepackage{tikz} 
\usetikzlibrary{calc}
\usepackage[export]{adjustbox}
\usepackage[normalem]{ulem}

\usepackage[inline, final]{showlabels}
\usepackage{tensor}

\usepackage{listings}
\usepackage{xcolor}

\definecolor{comment_red}{rgb}{0.5, 0, 0}
\makeatletter
\g@addto@macro\bfseries{\boldmath}  
\makeatother

\newcommand{\ba}{\begin{eqnarray}}
\newcommand{\ea}{\end{eqnarray}}

\newcommand{\vect}[1]{\boldsymbol{#1}}
\newcommand{\Msun}{M_\odot}
\newcommand{\td}{{\rm d}}
\newcommand{\be}{\begin{equation}}
\newcommand{\ee}{\end{equation}}
\newcommand{\bea}{\begin{equation} \begin{aligned}}
\newcommand{\eea}{\end{aligned} \end{equation}}

\newcommand{\Mc}{\mathcal{M}_c} 

\newcommand{\glow}{\texttt{GLoW}}

\definecolor{grey}{rgb}{0.4,0.4,0.4}
\definecolor{dullmagenta}{rgb}{0.4,0,0.4}
\definecolor{darkblue}{rgb}{0,0,0.4}
\definecolor{midblue}{rgb}{0,0,0.5}
\definecolor{midred}{rgb}{0.5,0,0}
\definecolor{orange}{rgb}{1,0.5,0}
\definecolor{lightbrown}{rgb}{0.75,0.5,0.25}
\definecolor{tan}{cmyk}{0.14,0.42,0.56,0}
\definecolor{djunglegreen}{cmyk}{0.99,0,0.52,0}
\definecolor{lightgreen}{rgb}{0,1,0}
\definecolor{olivegreen}{cmyk}{0.64,0,0.95,0.40}
\definecolor{midgreen}{rgb}{0.0,0.675,0.0}
\definecolor{darkgreen}{rgb}{0,0.5,0}
\definecolor{ceruleanblue}{rgb}{0.0, 0.2, 0.7}
\definecolor{burgundy}{rgb}{0.5, 0.0, 0.13}
\definecolor{hvred}{RGB}{186,12,47}

\hypersetup{
    colorlinks=true,
    linkcolor=ceruleanblue,
    filecolor=ceruleanblue,      
    urlcolor=midblue,
    citecolor=burgundy,
}

\usepackage[all]{xy}
\usepackage{amsfonts}
\makeatletter
\def\l@subsubsection#1#2{}
\makeatother

\begin{document}

\title{Signatures of $10-10^4\,{\rm M}_{\odot}$ Dark Matter halos in LISA via Stochastic Diffraction} 

\author{Han Gil Choi}
\email{hangil.choi@aei.mpg.de}
\affiliation{Max Planck Institute for Gravitational Physics (Albert Einstein Institute) \\
 Am Mühlenberg 1, D-14476 Potsdam-Golm, Germany}
 \affiliation{Cosmology, Gravity, and Astroparticle Physics Group, Center for Theoretical Physics of the Universe,
Institute for Basic Science (IBS), Daejeon, 34126, Korea}

\author{Juan Urrutia}
\email{juan.urrutia@kbfi.ee}
\affiliation{Keemilise ja Bioloogilise F\"u\"usika Instituut, R\"avala pst. 10, 10143 Tallinn, Estonia}
\affiliation{Department of Cybernetics, Tallinn University of Technology, Akadeemia tee 21, 12618 Tallinn, Estonia}

\author{Miguel Zumalac\'arregui}
\email{miguel.zumalacarregui@aei.mpg.de}
\affiliation{Max Planck Institute for Gravitational Physics (Albert Einstein Institute) \\
 Am Mühlenberg 1, D-14476 Potsdam-Golm, Germany}

\begin{abstract}
Cold Dark Matter predicts a population of low-mass halos which are sensitive to its fundamental nature and the primordial power spectrum, yet remain undetected. Although elusive, their discovery may be possible thanks to wave-optics lensing of gravitational waves (GWs) by the superposition of many halos along the line of sight. We study the statistical properties of \textit{stochastic diffractive lensing}, which imprints correlated fluctuations on the amplitude and phase of the original waveform. The stochastic distortions can be described by an orthogonal basis that captures the dominant ``tones'' associated with the dark matter properties, or \textit{dark timbre}, which is not degenerate with binary source parameters. LISA is most sensitive to halos of $\mathcal{O}(10\text{--}10^4\,M_\odot)$, and because the imprint recurs in every source, stacking $\sim(50,500)$ loud binaries could confirm them at the $(2,5)\sigma$ level ($\sim0.2$ to $\gtrsim4\sigma$ for realistic merger rates and different concentration estimations). The per-event signal is only $\mathcal{O}(10^{-3})$ in cold dark matter, demanding major advances in waveform accuracy and data analysis.  Even short of that reach, stochastic diffraction places stringent bounds on models that enhance small-scale structure, such as axion miniclusters and primordial black holes.
\end{abstract}

\maketitle

\noindent\textbf{Introduction  -- }\label{sec:intro}
A musical instrument sounds different when played in the open air, in a small room, or in a concert hall. The instrument produces the same notes, but the surrounding space reshapes their relative amplitudes and phases, imprinting its own acoustic signature on the sound. Gravitational-wave (GW) sources behave similarly: as their coherent signals propagate across the Universe, they diffract on the unresolved matter distribution along the line of sight. The resulting stochastic, frequency-dependent distortion, the Dark Timbre of the signal, is determined not only by the source, but also by the cosmic environment through which the wave has travelled. We will show that this imprint can reveal a population of low-mass dark matter (DM) halos that has so far remained beyond observational reach.

Cold dark matter (CDM) although being an extremely successful cosmological theory~\cite{Planck:2018vyg,1970ApJ...159..379R,1984ApJ...277..470P,Clowe:2006eq}, it predicts a population of DM halos down to planetary masses~\cite{Diemand:2005vz,Bullock:2017xww} that remains to be confirmed. These diffuse halos do not contain stars~\cite{Barkana:2000fd} and have extremely feeble gravitational interactions, which have prevented a robust detection below $\sim10^6 \,M_{\odot}$~\cite{Fairbairn:2022gar}. They are nevertheless a key prediction of CDM: their abundance, concentration and mass function are sensitive to the microphysics of dark matter~\cite{Bringmann:2009vf}, as well as the Universe's initial conditions~\cite{Bringmann:2025cht}. This has sparked an active search program, including signatures in strong lensing images, stellar wakes, dwarf-galaxy dynamics, Lyman-$\alpha$, pulsar timing array searches and spectral distortions~\cite{Vegetti:2012mc,Vegetti:2023mgp,Buschmann:2017ams,Graham:2024hah,Irsic:2017ixq,Ramani:2020hdo,Chluba:2012we,Hsueh:2019ynk,Mao:1997ek,Ibata:2001iv,Banik:2019smi,Bonaca:2018fek,2018MNRAS.473.2060J,DES:2020fxi,Viel:2013fqw,Rogers:2020ltq,Chluba:2012we,2011JCAP...07..025K,Croon:2020wpr}.

Wave-optics lensing of GWs opens an additional avenue into the existence and properties of light DM halos~\cite{1999PThPS.133..137N,Takahashi:2003ix,Tambalo:2022wlm,Fairbairn:2022xln}. Frequency-dependent diffraction patterns carry information about the matter distribution~\cite{Choi:2021bkx,Savastano:2023spl,Urrutia:2024pos} and can be observed on GWs thanks to their phase coherence. Moreover, accurate models for GWs from compact binary systems can be used to infer both the source and lens parameters~\cite{Wright_2022,cheung2024probingminihalolensesdiffracted,Goyal:2025eqo,Urrutia:2021qak}.
While the parameters of individual DM halos can in principle be recovered, studies find that the probability of distinctive lensing by individual halos is small~\cite{Fairbairn:2022xln,Brando:2024inp}, at least for concentrations expected in CDM theories. A detection of CDM may be possible by exploiting the collective signal produced by a large number of low-mass lenses~\cite{1986ApJ...307...30D,Macquart:2004sh,Takahashi:2005ug,Oguri:2020ldf,Urrutia:2024pos,Zumalacarregui:2024ocb,Kim:2025njb,Ando:2026eam,Ando:2026poq,Zumalacarregui:2026uqs,Amoruso:2026txw}.

In this \textit{letter}, we develop the formalism required to describe this imprint and assess the prospects for its detection as a probe of the light end of the halo mass function. We find that LISA~\cite{LISA:2017pwj} is the ideal detector to search for these signatures and that the effect is primarily driven by halos of $\mathcal{O}\left(10-10^4\right)\,M_{\odot}$. If more than $\sim(50\,,500)$ loud binaries are detected and waveform systematics~\cite{Yi:2025pxe} are lower than $\mathcal{O}(1\%)$, the existence of these DM halos could be confirmed at the $(2\,,5) \sigma$ CL. When contrasting this with estimations for the merger rates, we find substantial uncertainty in the statistical evidence, which can range from $\sim0.2\sigma$ to $\gtrsim4\sigma$ after 10 yr of observation. Furthermore, there is significant uncertainty due to the concentration of the halos~\cite{supplemental} (see Sec.~\ref{sec:concentration}) and their evolution with redshift. What is guaranteed is the ability to constrain modifications to CDM that enhance small-scale structure, such as axion miniclusters~\cite{Hogan:1988mp} or primordial black holes~\cite{Carr:1974nx}, shedding light on otherwise inaccessible scales. The stochastic fluctuations distort the waveform independently of its use as a DM probe, so it could be an unavoidable feature of GW propagation in a DM universe with relevant consequences for GW phenomenology broadly defined.

\noindent\textbf{Stochastic diffraction -- }\label{sec:swo}As a GW travels cosmological scales,  it is going to encounter many DM halos and structures. Under the thin-lens, Born, and Eikonal approximations~\cite{Takahashi:2003ix,1992grle.book.....S}, the amplification factor $F(f)$, which quantifies the effect of a lens, $\tilde{h}_{L}=F(f)\tilde{h}$, is given by
\begin{equation} \label{eq:multiF}
    F(f) = \prod_{j=1}^{N} \frac{w(f)}{ 2i\pi}\int d^2 x_j~e^{ i w(f)T_j(\bm{x}_{j},\bm{x}_{j+1})}\, ,
\end{equation}
where $\bm{x}_j = \bm{\xi}_j/\xi_0$, $j=1,2,\dots,N$ are dimensionless position vector on the $j$-th lens plane normalized with characteristic length scale $\xi_0$, which we define as the Fresnel scale at the minimal frequency $f_{\rm min}$ considered in the simulation, $\xi_0=r_f(f_{\rm min})=\sqrt{D_{\rm eff}/[2\pi(1+z_l)f_{\rm min}]}$, with $D_{\rm eff}=D_lD_{ls}/D_s$ the effective lens distance~\cite{Takahashi:2005ug}. In the linear regime, we can re-express this as a single effective lens potential, see the Supplemental Material~\cite{supplemental} (Sec.~\ref{sec:multiplane}) for a complete derivation, such as 
\be\label{eq:delta_F}
    \Delta F(w) \;=\; -\frac{w^{2}}{2\pi}\,
    \int \td^{2}\!x\;
    e^{\,iw\,|\vect{x}|^{2}/2}\Psi(\vect{x}),
\ee
where the lensing potential $\Psi(x)$ is the contribution from all the DM halos considered. Also, notice that the lensing potential has no symmetry, as is the usual case for isolated lenses. After performing a Fourier transformation, we can express it as
\bea\label{eq:delta_F_2}
    \Delta F(w) \;&=\; \int\frac{\td^{2}k}{(2\pi)^{2}}\,H(\vect{k};w)\,\hat{\kappa}(\vect{k})\, ,\\
    H(k,w) \;&\equiv\; \frac{2iw}{k^{2}}\bigl[e^{-ik^{2}/(2w)}-1\bigr]\, ,
\eea
notice that $k\equiv k_{\rm phys.}\times r_f(f_{\rm min})$ so it is dimensionless. For a transient GW source $\hat{\kappa}$ is a random field, and so is the diffractive fluctuation $\Delta F(w)$. It is convenient to split it into two contributions: a \emph{Gaussian} component, sourced by the many low-mass halos that populate each wave-optics volume, and a non-Gaussian \emph{Poisson} component from the rare, massive halos that produces the characteristic tail of weak lensing distributions. We characterise the field's second moments below, and defer the validity of this description to the Supplemental Material~\cite{supplemental} (Sec.~\ref{sec:validity}).
\begin{figure}
    \centering
    \includegraphics[width=0.8\columnwidth]{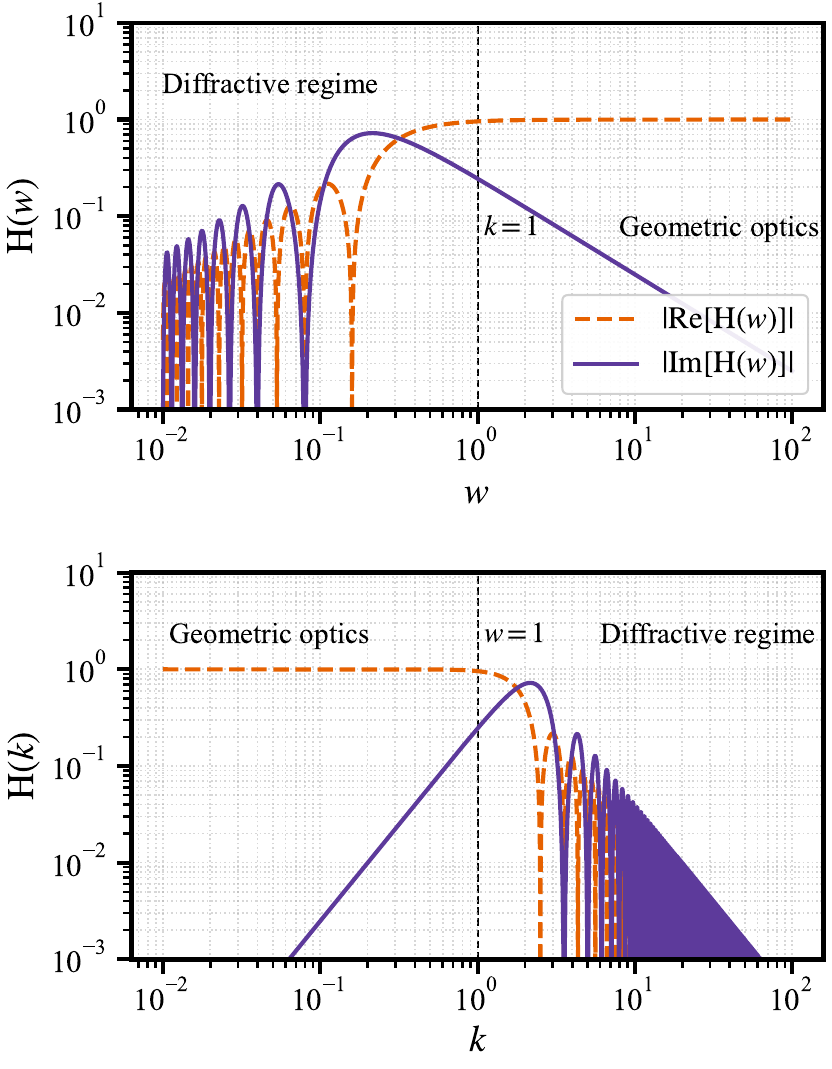}
    \caption{Structure of the windowing functions (Eq.~\ref{eq:delta_F_2}) \textit{Top panel:} real and imaginary components of $H(k,w)$ as a function of the dimensionless frequency $w$ for a mode with $k=1$. \textit{Bottom panel:} the same components at fixed $w=1$ as a function of the dimensionless wavenumber $k$.}
    \label{fig:window_func}
\end{figure} 
The statistics of $\hat\kappa$ are set, at the two-point level, by the convergence power spectrum,
\be\label{eq:pedag_Pkappa}
    \langle\hat\kappa(\vect{k})\,\hat\kappa^{*}(\vect{k}')\rangle
    \;=\; (2\pi)^{2}\,\delta^{(2)}(\vect{k}-\vect{k}')\,P_{\kappa}(k),
\ee
which defines the convergence power spectrum $P_{\kappa}(k)$.%
\footnote{Since $\kappa(\vect{x})$ is real, its Fourier transform satisfies $\hat\kappa(-\vect{k})=\hat\kappa^{*}(\vect{k})$,
so that
\be\label{eq:pedag_pseudoPkappa}
    \langle\hat\kappa(\vect{k})\,\hat\kappa(\vect{k}')\rangle
    \;=\; (2\pi)^{2}\,\delta^{(2)}(\vect{k}+\vect{k}')\,P_{\kappa}(k).
\ee
This second relation is what makes $\Delta F(w)$ a non-circular complex
Gaussian: the pseudo-covariance
$\langle\Delta F(w)\,\Delta F(w')\rangle$ does not vanish in general,
and the real and imaginary parts of $\Delta F$ are neither independent
nor identically distributed at fixed $w$, and although both two-point functions are needed to fully characterise the field, we have explicitly checked that considering only the covariance is sufficient to estimate the detectability.}
The central object for the detectability is the frequency covariance of the fluctuation, obtained by substituting Eq.~\eqref{eq:delta_F_2} into the two-point function,
\bea\label{eq:cov_def}
    C(w,w') &\;\equiv\; \bigl\langle\Delta F(w)\,\Delta F^{*}(w')\bigr\rangle
    \;\\
    &=\; \frac{1}{2\pi}\int_{0}^{\infty}\!k\,\td k\;
                   H(k;w)\,H^{*}(k;w')\,P_{\kappa}(k)\,.
\eea
The covariance $C(w,w')$ is a general second-moment quantity that does not assume Gaussianity; it is nonetheless all that is needed both to realise the Gaussian component of the field, through a Cholesky factorisation. Its diagonal gives the variance of the fluctuations,
\be\label{eq:pedag_variance}
    \bigl\langle|\Delta F(w)|^{2}\bigr\rangle
    \;=\; \frac{8w^{2}}{\pi}\int_{0}^{\infty}\!\frac{\td k}{k^{3}}\,
    P_{\kappa}(k)\,\sin^{2}\!\left(\frac{k^{2}}{4w}\right)\, .
\ee
The structure of the windowing functions is shown in Fig.~\ref{fig:window_func}. The top panel shows how contributions are suppressed in the \textit{diffractive limit} at low frequency, with the imaginary part dominating the correction. Conversely, in the \textit{geometric optics limit} at high frequencies, the amplification becomes frequency-independent and purely real; this region also corresponds to the single image regime. In the intermediate region, diffraction induces frequency-dependent modulations, with contributions to both the real and imaginary parts. The bottom panel shows the behaviour at a fixed dimensionless frequency. Low-$k$ modes correspond to scales larger than the Fresnel length and contribute to the geometric regime, as expected. Frequency-dependent, wave-optics corrections are associated to small scales (intermediate and high $k$). 

Universe homogeneity is reflected in the vanishing mean $\langle\Delta F(w)\rangle=0$, which follows from $\langle\hat\kappa(\vect{k})\rangle = 0$ for a mean-subtracted field; this is the most trivial of a series of relations fulfilled by $\Delta F$~\cite{Takahashi:2005ug,Nakazono:2026vei}. By the central limit theorem, the Gaussian component of $\Delta F(w)$ is realised when many halos contribute comparably to the variance, while non-Gaussian tails appear when few massive halos dominate, as is characteristic of one-point functions and weak-lensing distributions~\cite{Bartelmann:1999yn}.
\begin{figure}
    \centering
    \includegraphics[width=\columnwidth]{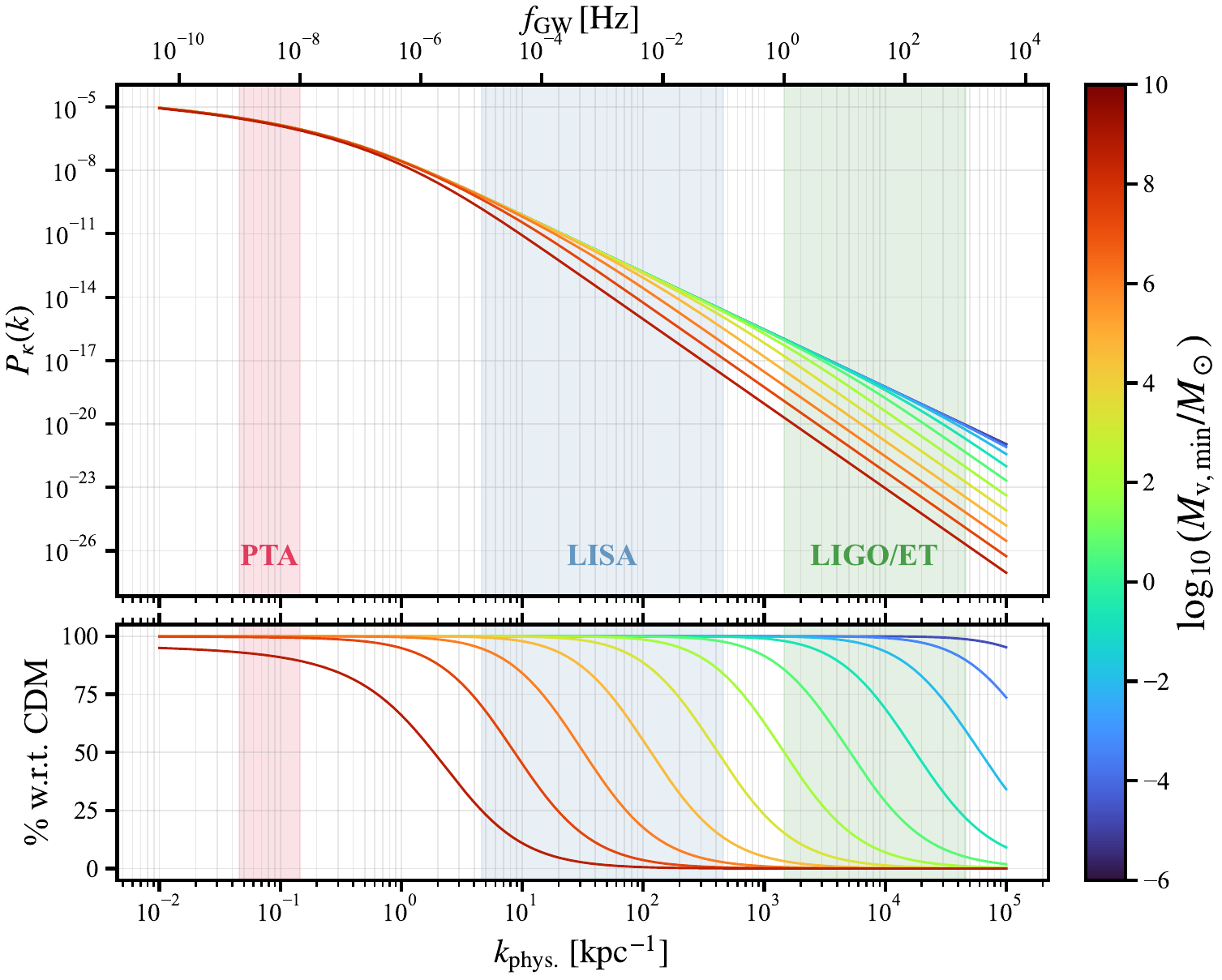}
    \caption{Convergence power spectrum in CDM.
    \textit{Top panel:} dimensionless convergence power spectrum for a source at $z_s=2$, as a function of frequency at $z_l=1$, and different cuts in the mass function. 
    \textit{Lower panel:} relative difference in the convergence power spectrum with respect to CDM.}
    \label{fig:Pk_var}
\end{figure}

We now specialise the convergence power spectrum to CDM, neglecting the bias term~\cite{Sheth:1999mn}, which can be expressed as
\bea\label{eq:Pkappa_1h_z}
    P_\kappa(k) &= \int_0^{z_s}\frac{\td z\left[(1+z)\,\,r_f(z,f_{\rm min})\right]^2}{H(z)}\int \td M\,\frac{\td n}{\td M}(M,z)\,\\
    &\times\bigl[\hat\kappa_{\rm NFW}(k;\kappa_0(M,z),b(M,z))\bigr]^2\,,
\eea
where $\td n/\td M$ is the halo mass function~\cite{Sheth:1999mn,1974ApJ...187..425P}, and $\hat\kappa_{\rm NFW}$ is the analytic Hankel transform of the projected NFW profile~\cite{Navarro:1996gj}, characterised by convergence normalisation $\kappa_0=\rho_s r_s/\Sigma_{\rm cr}$ and inverse scale radius $b=r_f(f_{\rm min})/r_s$, which depend on the halo mass and concentration. The concentration sets the scale radius at fixed halo mass, so the overall amplitude of $\hat\kappa_{\rm NFW}$ scales as $\kappa_0/b^2$ (up to a logarithm), while the per-halo variance $\propto\kappa_0^2$ grows as $c^4/\mu(c)^2$; the concentration is therefore a leading-order control on the strength of the signal at fixed halo abundance. The closed form of $\hat\kappa_{\rm NFW}$ in terms of the sine and cosine integrals, together with the impact of halo concentration, is given in the Supplemental Material~\cite{supplemental} (Secs.~\ref{sec:halo} and~\ref{sec:concentration}).
\begin{figure*}[th]
    \centering
    \includegraphics[width=0.9\textwidth]{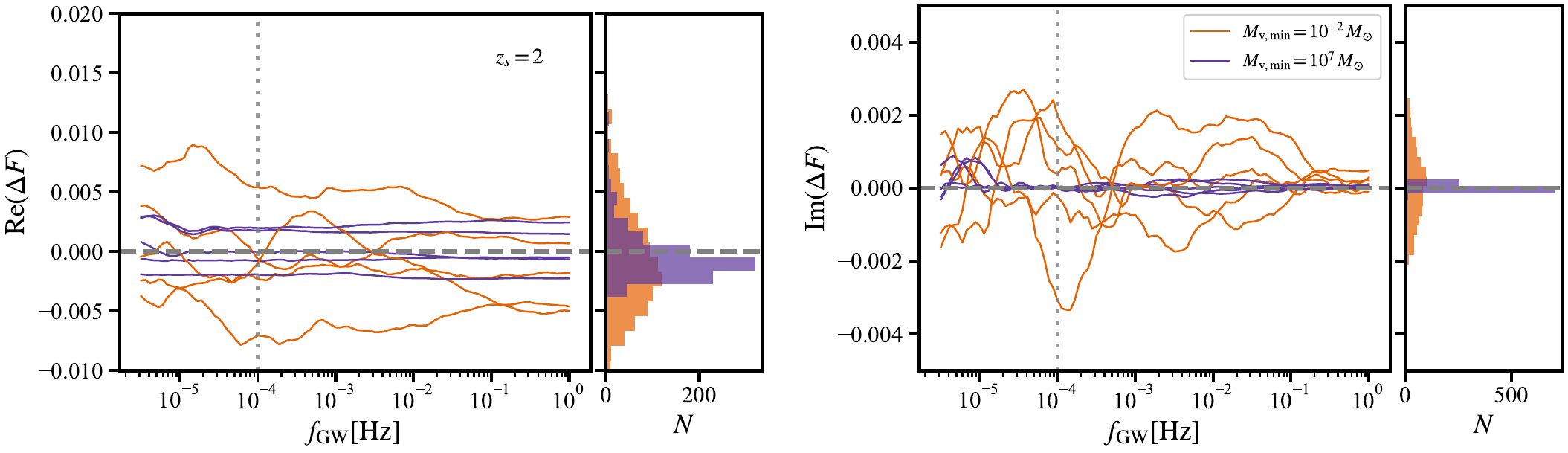}
    \caption{Realisations of the stochastic diffraction, with both the \textit{Gaussian} and \textit{Poisson} components, for a binary at $z_s=2$, simulated up to $10$ fresnel scales and for different cuts of the halo mass function. Blue lines corresponds to all halos above $1\,M_{\odot}$, orange only heavier halos above $10^7\,M_{\odot}$. The left/right panels shows the real/imaginary component of $\Delta F=F-1$. Including lighter halos increases the amplitude of frequency-dependent fluctuations, especially on the imaginary part (i.e. phase fluctuations).}
    \label{fig:CDM_fluctuations}
\end{figure*}

The top panel of Fig.~\ref{fig:Pk_var} shows the convergence power spectrum $P_\kappa(f)$ for the CDM halo mass function from~\cite{Urrutia:2025fvp} with different lower mass cuts. $P_\kappa$ decreases with frequency because higher $f$ probes smaller scales, which are more populated but contribute less per halo. We also show the approximate relevant frequency range for different types of detectors, including PTAs~\cite{NANOGrav:2023gor,EPTA:2023sfo}, LISA~\cite{LISA:2017pwj}, and ground-based detectors like LIGO-VIRGO-KAGRA~\cite{LIGOScientific:2026sit} and ET~\cite{ET:2025xjr}. In the right panel, we can see the relative difference of the convergence power spectrum with respect to the different halo mass cuts. The comparison identifies the range of halo masses each detector can probe: the in-band convergence power keeps growing as lighter halos are included until it saturates at $M_{\rm v}\sim10^{10}\,M_{\odot}$ for PTAs, $\sim1\,M_{\odot}$ for LISA and $\sim10^{-6}\,M_{\odot}$ for ground-based detectors, below which lighter halos add no in-band power. We leave the exploration of the stochastic diffraction in other detectors for future work.

With the convergence power spectrum in hand, we generate explicit realisations of the stochastic diffraction. We evaluate $\Delta F$ as a sum of two pieces split at a mass threshold $M_{\rm cut}$: a Gaussian piece for the numerous light halos ($M<M_{\rm cut}$), drawn from the covariance Eq.~\eqref{eq:cov_def} via a Cholesky factorisation, and an explicit Poisson piece for the rare heavy halos ($M>M_{\rm cut}$), sampled halo-by-halo through their single-halo contribution $\Delta F_1(w;M,r_c)$; both, together with the number of halos in the wave-optics volume, are detailed in the Supplemental Material~\cite{supplemental} (Sec.~\ref{sec:halo}). The Poisson piece carries a deterministic mean-field subtraction: since we sample only the overdense halos and not the compensating underdense regions, subtracting the expected halo contribution over the sampling disk plays the role of the unsimulated voids and restores the cosmic-mean reference $\langle\Delta F\rangle=0$ (Sec.~\ref{sec:halo}). The convergence integral Eq.~\eqref{eq:Pkappa_1h_z} runs over the full line of sight, so no single lens redshift enters the calculation; a reference lens redshift is needed only to map dimensionless wavenumbers to a physical scale, and hence to frequency, in the illustrative figures, for which we take $z_l=1$ and a source at $z_s=2$ (the convergence spectra of Fig.~\ref{fig:Pk_var} span the GW-relevant range $(10^{-10},\,10^{4})\,{\rm Hz}$). For the moment we adopt the CDM halo mass function, deferring modified matter power spectra to later.

We now have the framework to produce realistic realisations of the stochastic diffraction. Figure~\ref{fig:CDM_fluctuations} shows the resulting fluctuations for a source at $z_s=2$ and two lower mass cuts of the halo mass function, $M_{\rm v}=10^{-2}\,M_{\odot}$ and $M_{\rm v}=10^{7}\,M_{\odot}$, with halos simulated up to $M_{\rm v,max}=10^{8}\,M_{\odot}$, beyond which the halos host dwarf galaxies~\cite{2024MNRAS.535....1T}, and a different profile would be required~\cite{DELVE:2025ugp}. Because we extend the wave-optics volume to $N=10$ Fresnel scales, the highest-mass cut is dominated by halos that sit deep in the geometric-optics limit. When only halos above $10^7M_\odot$ are included, the imaginary part essentially vanishes and the amplification is nearly frequency-independent, particularly at high frequencies. Including halos down to $10^{-2}\,M_{\odot}$ changes the picture qualitatively: the real part develops strongly frequency-dependent fluctuations, and the imaginary part, which is sourced by modes still in the diffractive regime, becomes the most frequency-sensitive piece of the signal. The non-Gaussian tail is pronounced in the real part and negligible in the imaginary part. The reason is that non-Gaussianity is mainly due to rare heavy halos at large impact parameters, which contribute almost exclusively to the geometric, real-valued sector. As we will show below, the geometric contribution does not enter the test statistic and is, in any case, dominated by ordinary weak-lensing distortions from larger structures; this justifies treating the detectable signal in the Gaussian approximation throughout. 
\begin{figure*}
    \centering
    \includegraphics[width=\textwidth]{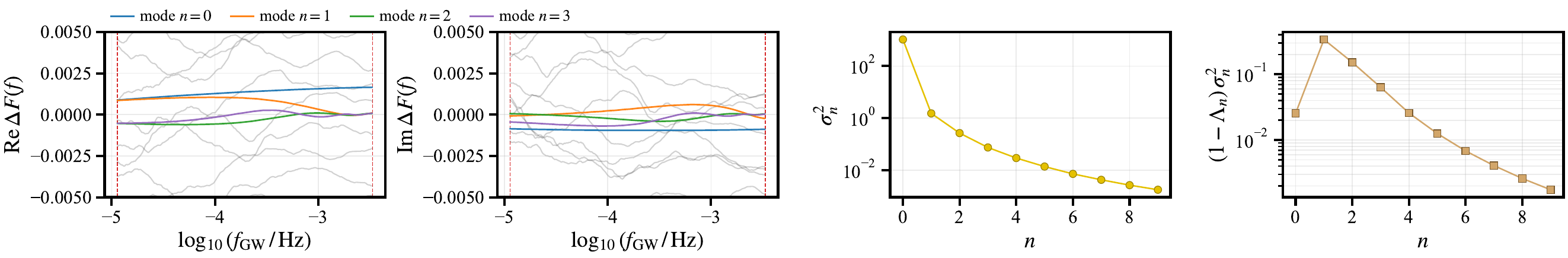}
    \caption{
    \textit{Left panels:} Real and imaginary parts of stochastic diffraction realisations for a source at $z_s=2$ (gray) and the first 4 Dark-Timbre harmonics (colors)
    \textit{Right panels:} Each of the harmonics is uncorrelated and characterised only via its standard deviations, shown in yellow. In the rightmost panel, we see the relative power of each mode when accounting for degeneracies with binary parameters: note that the n=0 mode is highly degenerate with the luminosity distance. The binary has $\mathcal{M}_c=10^6\,,\eta=0.25$ and is observed by LISA during $5$ year before merger.}
    \label{fig:spectral_decomp}
\end{figure*}

\noindent\textbf{The Dark Timbre and its Detectability -- } Stochastic diffractive lensing is not an entirely random process: it encodes the properties of DM in the statistical properties of amplification fluctuations, including correlations across different frequencies.
In order to assess the discovery potential and interpretation, it is essential to ``order'' the contributions and assess their degeneracy with intrinsic binary parameters. We will focus on LISA observations of massive black holes. We will define the inner product in the space of amplification factor fluctuations $\Delta F(f)$~\cite{Finn:1992xs,Poisson:1995ef}
\bea
    \langle u, v\rangle_W
    &\;\equiv\; 4\int df\; W(f)\, u^{*}(f)\,v(f)\, ,\\
    W(f) \;&\equiv\; \frac{|\tilde{h}(f)|^{2}}{S_n(f)}\, ,
\label{eq:inner}
\eea
where $u,v$ represent amplification factors. The weight $W$ is optimized to the unlensed frequency-domain waveform $\tilde h$ and the LISA detector noise $S_n(f)$, including unresolved sources~\cite{Lewicki:2021kmu}. We construct the Karhunen--Lo\`eve~\cite{2015arXiv150907526A} expansion of the covariance $C(f,f')$ weighted by this inner product (see Sec.~\ref{sec:detstat} in Supplemental Material). The result gives an orthonormal basis of eigenvectors $\psi_n$ that can be used to describe stochastic diffraction fluctuations caused by DM. This construction also gives the probability density function of the coefficients, which are uncorrelated by construction. We will refer to these coefficients as the Dark Timbre, using the analogy with sound where different harmonics are modified. The resulting representation of the stochastic diffraction in the Dark Timbre basis is 
\begin{equation}
    \Delta F(f)
    \;=\; \sum_n c_n\,\psi_n(f)\,,
    \qquad
    c_n\sim \mathcal{CN}\bigl(0,\sigma_n^{2}\bigr)\, ,
\label{eq:KL}
\end{equation}
where $\mathcal{CN}$ refers to an uncorrelated complex Gaussian from where the Dark Timbre coefficients are drawn; here $\sigma_n$ are real, and $\psi_n(f)$ are, in general, complex functions. As discussed above, we focus on the Gaussian limit of the signal dominated by many light halos; we are neglecting the Poisson contribution to the signal, since this mode is degenerate with the $(n=0)$ and does not contribute to the detectability. 

To address the degeneracy with the binary properties, let us consider the effect of a small change in the parameter vector $\vect{\theta}=(\mathcal{M}_c,\eta,\chi_1,\chi_2,t_c,\phi_c,d_L,\iota,\ldots)$, such as the waveform is modified by
\be
    \delta_a(f) \;\equiv\;
    \partial_a \tilde{h}(f,\vect{\theta})\big|_{\theta_0}\, ,
\label{eq:param_directions}
\ee
where $a$ labels binary parameters. Following the Cutler-Vallisneri formalism~\cite{Cutler:2007mi}, the orthogonal projection of $\psi_n$ onto the parameter subspace yields the leakage in terms of
\be\label{eq:alpha_def}
    \alpha_{an}\;\equiv\;\langle\psi_n,\delta_a\rangle_w\, ,
    \qquad
    \Gamma_{ab}\;\equiv\;\mathrm{Re}\,\langle\delta_a,\delta_b\rangle_w\, ,
\ee
where $\Gamma_{ab}$ is the Fisher matrix of the unlensed waveform~\cite{Poisson:1995ef}; the real part appears because the binary parameters $\vect{\theta}$ are real, so only the real part of the product enters the parameter metric. The leakage of mode $n$ into the binary parameters is then
\be\label{eq:Lambda}
    \Lambda_n \;\equiv\; \sum_{a,b}\mathrm{Re}\bigl[\alpha_{an}^{*}\bigr]\,(\Gamma^{-1})^{ab}\,\mathrm{Re}\bigl[\alpha_{bn}\bigr]\, ,
\ee
which is an orthogonal projection of mode $n$ onto the real parameter subspace, i.e.\ the fraction of the mode's power that is degenerate with the binary parameters and thus reabsorbed by a shift in $\vect{\theta}$.
The constraint that the unlensed waveform places on $\vect{\theta}$ is already encoded in $\Gamma_{ab}$, so Eq.~\eqref{eq:Lambda} is the proper full-flat-prior marginalisation over all binary parameters. It follows that the Dark Timbre detection average significance for one event is
\be\label{eq:chi2}
    \langle\Delta\chi^{2}\rangle
    \;=\; \sum_{n}\sigma_n^{2}\bigl(1-\Lambda_n\bigr)\, ,
\ee
where $\Lambda_n$ accounts for degeneracies with binary parameters.
Concretely, the detection tests the null hypothesis $\mathcal{H}_0$ (no lensing), under which the whitened Dark-Timbre coefficients of the residual relative to the maximum-likelihood unlensed waveform $h_0(\vect{\theta}_0)$ are normalized noise, $d_n\sim\mathcal{N}(0,1)$, against $\mathcal{H}_1$ (lensing), $d_n\sim\mathcal{N}\!\left(0,\,1+\sigma_n^2(1-\Lambda_n)\right)$
[see~\cite{supplemental}, Sec.~\ref{sec:detstat}].

As we will show, no binary is loud enough in LISA to detect the Dark Timbre with sufficient confidence. Since it is going to be present in every source, it is possible to account for the combined evidence from all the detected binaries in a given catalogue. We can add the evidence from each event by 
\be\label{eq:sigmaEvi}
\sigma^2_{\rm Evi.}=\sum_i\rho_i^2=\frac{1}{4}\sum_{i,n}\left[\sigma_{n,i}^{2}\left(1-\Lambda_{n,i}\right)\right]^2 \, , 
\ee
where $i$ labels the detectable binaries and $\rho_i$ is the per-event detection significance derived in~\cite{supplemental} (Sec.~\ref{sec:detstat}). For the waveform, we will use an inspiral-merger-ringdown template~\cite{Ajith:2007kx} averaged over sky orientations~\cite{Gerosa:2019dbe,Finn:1992xs} and with a randomly assigned time to merger. \footnote{We also consider more realistic waveforms with precessing spins: these more detailed models lead to comparable detectability (within a factor of $2$) and sometimes even enhanced; see Sec.~\ref{sec:spin} for details. Furthermore, we consider including higher modes and the detectability is significantly increased, reducing the number of binaries necessary to get the same evidence by $\sim50\%$, see Sec.~\ref{sec:spin} for a comprehensive discussion.}

Fig.~\ref{fig:spectral_decomp} shows how the stochastic diffraction gets projected onto Dark Timbre harmonics for an equal mass, non-spinning binary with $\mathcal{M}_c=10^6$ at $z_s=2$ seen by LISA during $5$ years before merger. Note that the harmonics $\psi_n$ are weighted by $W(f)$, which peaks around the merger, giving more weight to higher frequencies in this realisation. We can see in the centre-right plot the power of each mode; the most power is in the $n=0$ mode, and then it has a power-law decay. When considering the power of each mode after the degeneracies with the parameter estimation, we can see how the $n=0$ is very suppressed since its slow-varying nature can be mimicked by biasing the parameters. In this case, the majority of the power is in the $n=[1,4]$ (especially $n=1$). For this case, we find that $\langle\Delta\chi^2\rangle=0.64$, corresponding to a per-event significance $\rho\simeq0.12$, so that $\sim110$ binaries will be necessary to detect this effect at the $2\sigma$ C.L.

\noindent\textbf{Bounds on Dark Halos -- }
To explore how we can use stochastic diffraction to constrain DM models beyond CDM, we will consider the matter power spectrum $\mathcal{P}(k)$, which we generate using the method from~\cite{Eisenstein:1997ik} and adjusted to the Planck results~\cite{Planck:2018vyg}. We will consider changes to the power spectrum of the following shape
\bea
&\mathcal{P}(k;\,k_0,k_{\rm cut}) = \mathcal{P}_{\rm CDM}(k)\,\Theta(k_0-k)\\
&+ \mathcal{P}_{\rm CDM}(k_0)\,\Theta(k_{\rm cut}-k)\,\Theta(k-k_0)\, .
\eea
with $k_{\rm cut}\geq k_0$. This form allows for an enhancement ($P_k={\rm const}$) between $k_0\leq k<k_{\rm cut}$ and a drop for $k>k_{\rm cut}$, providing a simple parameterisation valid for models that enhance or suppress small-scale structure, see Sec.~\ref{sec:beyondCDM}. The matter power spectrum enters the computation through the halo mass function, which we computed following Ref.~\cite{Urrutia:2025fvp}. Notice also that the concentration of the halos is adjusted accordingly, since the prescription from~\cite{2016MNRAS.456.3068O} accounts for variations in the variance of perturbations, see~\cite{supplemental} (Sec.~\ref{sec:concentration}) for a more complete discussion. Many of the models we will consider further enhance the concentration, i.e. Fuzzy DM predicts a core~\cite{Chan:2021bja} due to the wave nature of DM, so our benchmarks should be read as conservative estimates.

The minimum-mass cutoff explored above is equivalent to setting $k_{\rm cut}=k_0(M_{\rm v})$, where $k_0(M_{\rm v})=\left(4\pi\bar\rho_{\rm m,0}/3M_{\rm v}\right)^{1/3}$ is the comoving wavenumber enclosing the mass $M_{\rm v}$ at the cosmic mean density. This helped us identify the most relevant mass scales contributing to the signal, and provides a reasonable approximation to DM models with a UV cut-off, such as Fuzzy DM~\cite{Marsh:2015xka} or Warm DM~\cite{Bode:2000gq}. On the other hand, DM models with a scale-invariant enhanced power-spectrum, like primordial black holes~\cite{Inman:2019wvr,Carr:2018rid} or axion miniclusters~\cite{Kolb:1993zz,Fairbairn:2017sil,OHare:2021zrq}, can be approximated with models with $k_{\rm cut}>k_0$. More details on how to map physical properties of these DM models to the $(k_{\rm cut}\,, k_0)$ parametrisation can be found in~\cite{supplemental} (Sec.~\ref{sec:beyondCDM}). 

\begin{figure}
    \centering
    \includegraphics[width=0.9\columnwidth]{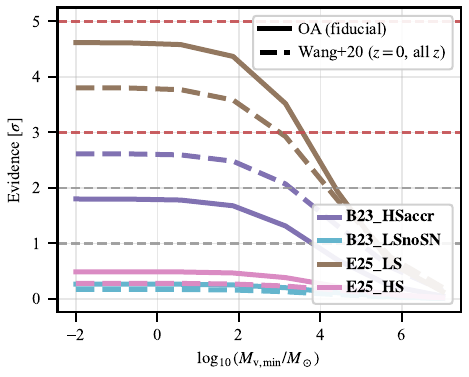}\\
     \includegraphics[width=0.9\columnwidth]{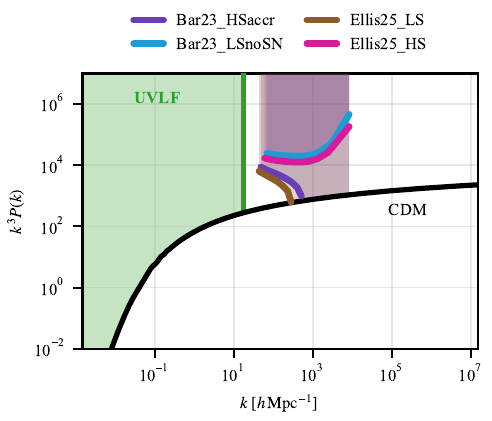}
    \caption{\textit{Top panel:} Evidence after $10$ years of observation with LISA, for the fiducial concentration relation but also using~\cite{Wang:2019ftp},  (dashed, [solid] resp) as a function of the minimum halo mass for different merger rates (colours). \textit{Lower panel:} Constraints on enhancements to the matter power-spectrum for the different merger rates considered in this work, as a reference, we show the constraints from the UV-luminosity function (UVLF) using JWST results~\cite{Urrutia:2025fvp}.}
    \label{fig:CDM_post}
\end{figure}

Fig.~\ref{fig:CDM_post} shows the evidence $[\sigma]$ after $10$ years of observation with LISA for different merger rates~\cite{Barausse:2020mdt,Barausse:2012fy,Klein:2015hvg,Antonini:2015sza,Barausse:2023yrx,Ellis:2025dpw}, for the presence of CDM halos up to different cuts in the halo mass function. The top panel of Fig.~\ref{fig:CDM_post} shows that the evidence can range from $\sim0.2$ to more than $4\sigma$ depending on the merger rates. We also considered the concentration relation from~\cite{Wang:2019ftp}, which is based on simulations and is not as optimistic as our fiducial. Since the results are reported only for $z=0$ we have kept this relation independent of redshift. We can see how, as expected, it weakens the results except for those merger rates which rely on more high-$z$ sources, our fiducial relation predicts from $z\sim2$ less concentrated halos than the one from~\cite{Wang:2019ftp}. A more comprehensive comparison with other concentration relations is given in \cite{supplemental} (see Sec.~\ref{sec:concentration}). Since the evidence is added quadratically, low-evidence events contribute negligibly, so the most important quantity is the number of strong events are realised. As anticipated by Fig.~\ref{fig:Pk_var}, LISA's sensitivity saturates around $\sim1\,M_{\odot}$ with no subsequent gain by including halos with lower masses. The lower panel in Fig.~\ref{fig:CDM_post} shows the expected limits on DM models that enhance the matter power spectrum on small scales, where the shaded regions represent values of $k_0>k_{\rm cut}$ that can be excluded in (10yr). Source populations that do not predict enough binaries to discover the CDM signal will still produce competitive constraints on scales which are inaccessible otherwise. The constraints on enhanced $P(k)$ are strongest in the population models with enough significance to discover standard CDM halos.

\begin{table}
\centering
\caption{Summary of constraints for enhancements of the matter power
spectrum, assuming a null detection of the dark timbre after
$T_{\rm obs}=10$\,yr at $2\sigma$.}
\label{tab:results}
\small
\setlength{\tabcolsep}{3pt}
\begin{tabular}{|l|c|c|c|}
\hline
 & $m_{\rm FDM}$ & $m_{\rm AMC}$ & $f_{\rm PBH}M_{\rm PBH}$\\
Model & $[{\rm eV}]$ & $[{\rm eV}]$ & $[M_\odot]$\\
\hline
\textbf{Ellis25 LS} & & & \\
\textbf{Bar23 HSaccr} & $\gtrsim 4\times10^{-17}$ & $\gtrsim 5\times10^{-13}$ & $\lesssim 1.5\times10^{-4}$\\
$k_0\lesssim 5.5\,{\rm kpc^{-1}}$ & & & \\
\hline
\textbf{Bar23 LSnoSN} & & & \\
$k_0\lesssim 0.8\,{\rm kpc^{-1}}$ & --- & $\gtrsim 2\times10^{-14}$ & $\lesssim 2\times10^{-2}$\\
\hline
\textbf{Ellis25 HS} & & & \\
$k_0\lesssim 1\,{\rm kpc^{-1}}$ & --- & $\gtrsim 3\times10^{-14}$ & $\lesssim 10^{-2}$\\
\hline
\end{tabular}
\end{table}

The limits on $P(k)$ can be translated into properties of dark matter. Tab.~\ref{tab:results} maps the $2\sigma$ constraints to Fuzzy Dark Matter, Axion Minicluster and Primordial Black Holes parameters. This illustrates how beyond-CDM theories can be probed by stochastic diffraction.

\noindent\textbf{Conclusions -- } We have developed the formalism for \textit{stochastic diffractive lensing} by the unresolved population of low-mass halos along the line of sight. Rather than searching for the rare signature of an individual lens, our approach exploits the correlated amplitude and phase fluctuations generated collectively by the halo population. These fluctuations are determined by the convergence power spectrum and leave an imprint on every propagating signal, with amplitude and frequency dependence fixed by the underlying theory. Decomposing these fluctuations into eigenmodes, the \textit{Dark Timbre}, organises the signal into a set of statistically independent components. Projecting these modes away from the binary-parameter subspace provides a direct measure of the observable signal and a practical route towards its extraction from GW data.

For massive black-hole binaries in the LISA band, the effect is dominated by halos of $\mathcal{O}(10\text{--}10^{4}),M_{\odot}$. The standard CDM signal is too weak to establish in any single event, but it is strong enough to be extracted at the population level. In the most favourable region of parameter space, a catalogue of $\sim(50,500)$ strong sources can confirm the existence of light CDM halos at $(2,,5)\sigma$ significance after ten years.
Three unknowns control this number: the underlying source population (cf.~Fig.~\ref{fig:CDM_post}); the halo concentration, uncertain enough to suppress the signal by an order of magnitude; and waveform systematics, which need to be controlled at the $\mathcal{O}(10^{-3})$ level or better.

Even if standard CDM cannot be detected, stochastic diffraction provides strong constraints on scenarios that enhance small-scale structure, including axion miniclusters and primordial black holes. Independent of its use as a CDM probe, this distortion is an unavoidable feature of GW propagation in any structured universe, and precision GW science will have to characterise, marginalise over, or actively model it regardless of one's prior on the halo mass function. More generally, the effect may bias astrophysical inference and propagate to other sources. In this picture, each binary acts as an individual instrument, while the global fit offers the possibility of extracting the universe's Dark Timbre from the full GW orchestra.

\vspace{5pt}\noindent\emph{Acknowledgements --}  We are thankful to Shin'ichiro Ando, Ville Vaskonen, Yann Gouttenoire, Jose Maria Ezquiaga, Hector Villarrubia-Rojo, Xikai Shan and Srashti Goyal for valuable discussions and comments on the draft. The work of JU was supported by the Estonian Research Council grants PRG803, PSG869, RVTT3 and RVTT7 and the Centre of Excellence program TK202. HGC was supported by IBS under the project code, IBS-R018-D3, and acknowledges financial assistance from the Max Planck Society Supporting Members. MZ is supported by the ERC Consolidator Grant GLOW (101230608). Views and opinions expressed are however, those of the author(s) only and do not necessarily reflect those of the European Union or the European Research Council Executive Agency. Neither the European Union nor the granting authority can be held responsible for them.
\bibliographystyle{apsrev4-2}
\bibliography{gw_lensing}

\newpage
\clearpage
\onecolumngrid

\begin{center}
\textbf{\large Signatures of $10-10^4\,M_{\odot}$ Dark Matter halos in LISA via Stochastic Diffraction} \\ 
\vspace{0.06in}{ Han Gil Choi, Juan Urrutia and Miguel Zumalacárregui} \\ 
\vspace{0.1in}
{SUPPLEMENTAL MATERIAL}
\vspace{0.1in}
\end{center}

\setcounter{equation}{0}
\setcounter{figure}{0}
\setcounter{section}{0}
\setcounter{table}{0}
\setcounter{page}{1}
\makeatletter
\renewcommand{\theequation}{S\arabic{equation}}
\renewcommand{\thefigure}{S\arabic{figure}}
\renewcommand{\thetable}{S\arabic{table}}
\renewcommand{\theHequation}{S\arabic{equation}}
\renewcommand{\theHfigure}{S\arabic{figure}}
\renewcommand{\theHtable}{S\arabic{table}}
\renewcommand{\theHsection}{S\arabic{section}}

\section{Multiplane lensing to effective single plane lensing}\label{sec:multiplane}

We consider the lensing effect of lens planes located at redshifts $z_j$, $j=1,2,\dots,N$ , where $z_1<z_2<\cdots<z_N$. The corresponding angular diameter distance is denoted as $D_j$. Let us denote $j=0$ and $j=N+1$ as indexes for observer and source, respectively~\cite{1986ApJ...310..568B,1992grle.book.....S,Feldbrugge:2019fjs,Ramesh:2021nnl,Jow:2022pux,Feldbrugge:2020ycp}. Under this notation, $z_{N+1}= z_s$ is a source redshift and the corresponding angular diameter distance is denoted as $D_{N+1} = D_s$. Also, angular diameter distance between the redshifts $z_j$ and $z_k$ ($j<k$) are denoted as $D_{jk} = D_k - D_j(1 + z_j)/(1+z_k)$, computed for a flat universe. The multiplane lensing amplification factor is given by
\begin{equation} \label{eq:multiF}
    F(f) = \prod_{j=1}^{N} \frac{w_{jj+1}(f)}{ 2i\pi}\int d^2 x_j~e^{ i w_{jj+1}(f)T_j(\bm{x}_{j},\bm{x}_{j+1})}
\end{equation}
where $\bm{x}_j = \bm{\xi}_j/\xi_0$, $j=1,2,\dots,N$ are dimensionless position vector on the $j$-th lens plane normalized with characteristic length scale $\xi_0$. We set the source position $\bm{x}_{N+1}\equiv \bm{x}_s=0$ without loss of generality. We define dimensionless frequency parameter $w_{jk}$ as
\begin{equation}
    w_{jk}(f) = \frac{(1+z_j)}{D_{\rm eff}^{jk}}2\pi f\xi_0^2
\end{equation}
where $D_{\rm eff}^{jk} \equiv D_j D_{jk}/D_{k}$, which is a generalisation of the effective distance of the single lens plane case. The time-delay functions on each lens plane $T_j$ are  
\begin{equation}
    T_j (\bm{x}_j,\bm{x}_{j+1})=  \frac{1}{2}\left|\bm{x}_j-\frac{D_j}{D_{j+1}}\bm{x}_{j+1}\right|^2 - \psi_{jj+1}(\bm{x}_j)
\end{equation}
Mass distribution of each lens plane is encoded in the lens potential $\psi_{jk}$ which is related to the lensing convergence $\kappa_{jk}$ by $\nabla^2_{\bm{x}_j}\psi_{jk} = 2 \kappa_{jk} = 8 \pi D_{\rm eff}^{jk}\Sigma_j$.

In the weak lensing regime, where $\psi_{jk}$ has a minor contribution to $F(f)$, the multi-plane lensing allows an effective single lensing description. To proceed, the following relations between the geometric contributions to the time delay is useful to simplify the notation 
\begin{equation}
\begin{split}
    & G_0(\bm{x}_j,\bm{x}_{k}) = \frac{w_{jk}}{2i \pi}e^{i w_{jk}|\bm{x}_j - \bm{x}_kD_j/D_k |^2/2}\\
    &\int d^2 x_{j}G_0(\bm{x}_j,\bm{x}_{k})  = 1~,\\
    &\int d^2 x_{k}G_0(\bm{x}_j,\bm{x}_{k}) G_0(\bm{x}_{k},\bm{x}_{l}) = G_0(\bm{x}_{j},\bm{x}_{l})~
\end{split}    ~.
\end{equation}
Then expanding perturbative on the lensing potential, $e^{-i w_{jj+1} \psi_{jj+1}}\simeq 1 - i w_{jj+1} \psi_{jj+1} $, Eq.~\eqref{eq:multiF}  becomes
\begin{equation}\label{eq:multiFlinear1}
\begin{split}
    F(f) &\simeq  \prod_{j=1}^{N} \int d^2 x_j~G_0(\bm{x}_j,\bm{x}_{j+1})(1 - i w_{jj+1}\psi_j(\bm{x}_j))\\
    & = 1 - i\sum_{j=1}^{N}\int d^2 x_j ~ G_0(\bm{x}_j,\bm{x}_{N+1})w_{jj+1}\psi_{jN+1}(\bm{x}_j) + \mathcal{O}(\psi^2)\\
    &\simeq 1 - \sum_{j=1}^{N}\frac{w_{jN+1}^2}{2\pi}\int d^2 x_j ~ e^{i w_{jN+1}|\bm{x}_j|^2/2}\psi_{jN+1}(\bm{x}_j) 
\end{split}    
\end{equation}
At the last line, we used the fact that $w_{jk}\psi_{jk} =w_{jk'}\psi_{jk'}$ for any $k'>k$. At this point, $F(f)$ described with Eq.~\eqref{eq:multiFlinear1} is nothing but sum of lensing effects coming from each lens planes which is as expected in linear regime where couplings between different lens planes are neglected. Now we construct its effective single lens plane description located at the redshift $z_l$. In Eq.~\eqref{eq:multiFlinear1}, one can rescale the dummy variable $\bm{x}_j$ as  $\bm{x} =\chi_j \bm{x}_j$ with the scale factor 
\be\label{eq:chij}
\chi_j\equiv \sqrt{\frac{(1+z_j)D_{\rm eff}}{(1+z_l)D^{jN+1}_{\rm eff}}}\, .
\ee
Here, $D_{\rm eff} = D_{l}D_{ls}/D_s$ is effective distance of the fiducial lens plane at $z_l$. Then the equation can be factorized with common phase factor and be reduced to
\begin{equation}\label{eq:multiFlinear2}
    F(f) \simeq 1 - \frac{w^2}{2\pi}\int d^2x~e^{iw|\bm{x}|^2/2}\Psi(\bm{x})~,
\end{equation}
where we introduce the effective lensing potential
\begin{equation}\label{eq:Psieff}
    \Psi(\bm{x}) = \sum_{j=1}^N \chi_j^2\psi_{jN+1}(\bm{x}/\chi_j)
\end{equation}
and fiducial dimensionless frequency
\begin{equation}
    w(f) = \frac{(1+z_l)}{D_{\rm eff}}2\pi f\xi_0^2
\end{equation}
In this work, we mainly consider the quantity
\begin{equation}\label{eq:multiFlinear3}
    \Delta F(f) =F(f)-1\simeq - \frac{w^2}{2\pi}\int d^2x~e^{iw|\bm{x}|^2/2}\left[\Psi(\bm{x})-\Psi(\bm{0})\right]~,
\end{equation}
where we subtracted the time-shift component $i w \Psi(\bm{0})$ so that it matches with the convention of the single-plane lensing $F(f)$~\cite{Nakamura:1997sw,1999PThPS.133..137N}.

The derivation above keeps the characteristic length $\xi_0$ a single global constant, so different lens planes carry different per-plane frequencies $w_{jN+1}$ and the rescaling factors $\chi_j$ of Eq.~\eqref{eq:chij} are non-trivial. We choose a scaling, such that the dimensionless frequency on every plane coincides with that of the fiducial plane. The natural choice is the Fresnel scale evaluated at $z_j$,
\begin{equation}\label{eq:xi0z}
    \xi_0(z_j) \;\equiv\; r_f(z_j,z_s,f_{\min}) \;=\; \sqrt{\frac{D^{jN+1}_{\rm eff}}{2\pi f_{\min}(1+z_j)}}~,
\end{equation}
where $f_{\min}$ is the lowest GW frequency considered. Substituting Eq.~\eqref{eq:xi0z} into the per-plane $w_{jN+1}$ gives, for every $z_j$,
\begin{equation}
    w_{jN+1}(f) = \frac{(1+z_j)}{D^{jN+1}_{\rm eff}}\,2\pi f\,\xi_0(z_j)^2 = \frac{f}{f_{\min}}\equiv w(f)~,
\end{equation}
so the dimensionless frequency is universal and the rescaling of Eq.~\eqref{eq:chij} becomes the identity, $\chi_j=1$. The dimensionless impact parameter and inverse scale radius of each halo are then $\bm x_j=\bm\xi_j/\xi_0(z_j)$ and $b_j=\xi_0(z_j)/r_s(M,z_j)$, both naturally evaluated in the halo's own slice. The redshift weighting that the global-$\xi_0$ formulation carries through $\chi_j^2$ in Eq.~\eqref{eq:Psieff} is now absorbed into $\xi_0(z_j)$ and the line-of-sight measure: the number of halos per unit dimensionless area on plane $j$ per unit redshift is
\begin{equation}\label{eq:Wlos}
    \frac{dN}{d^2x\,dz_j\,d\ln M} \;=\;\frac{ [\xi_0(z_j)\,(1+z_j)]^2}{H(z_j)}\,\frac{dn}{d\ln M}~,
\end{equation}
where $dn/d\ln M$ is the comoving halo mass function. The factor $(1+z_j)^2$ collects the proper-volume conversion of the comoving density and the proper line-of-sight depth, while $\xi_0(z_j)^2$ converts proper area on the lens plane into dimensionless area. The same measure fixes the mean convergence accumulated by the halo population per unit redshift,
\be\label{eq:kappahat}
    \frac{\td\hat\kappa}{\td z}
    \;=\;\frac{(1+z)^2}{H(z)\,\Sigma_{\rm cr}(z,z_s)}\int \td\ln M\;\frac{\td n}{\td\ln M}\,M\,,
\ee
using $\hat\kappa_{\rm NFW}(0)=M/(\Sigma_{\rm cr}\,\xi_0^2)$ for a profile truncated at the virial radius; it sets the deterministic mean-field term of Sec.~\ref{sec:halo}.

\newpage
\section{Convergence power spectrum and numerical realisation}\label{sec:halo}

The dimensionless convergence power spectrum quoted in the main text, Eq.~\eqref{eq:Pkappa_1h_z}, is built from the analytic Hankel transform of the projected NFW profile. Characterising each halo by the convergence normalisation $\kappa_0=\rho_s r_s/\Sigma_{\rm cr}$ and the inverse scale radius $b=r_f(f_{\rm min})/r_s$, the convergence admits the closed form
\bea\label{eq:kappa_hat_NFW}
    &\hat\kappa_{\rm NFW}(q;\kappa_0,b) = \frac{2\pi\kappa_0}{b^2}\\
    &\times\Bigl[
        {-}2\cos\!\Bigl(\frac{q}{b}\Bigr)\,\mathrm{Ci}\!\Bigl(\frac{q}{b}\Bigr)
        + \sin\!\Bigl(\frac{q}{b}\Bigr)
          \Bigl(\pi - 2\,\mathrm{Si}\!\Bigl(\frac{q}{b}\Bigr)\Bigr)
    \Bigr],
\eea
where $\mathrm{Si}$ and $\mathrm{Ci}$ are the sine and cosine integrals, and $q/b$ is the wavenumber in units of the inverse scale radius.

By the central limit theorem, $\Delta F(w)$ approaches a complex Gaussian when many halos contribute comparably to the variance; when instead the contributing halos are few, the massive ones dominate and non-Gaussian tails appear, as is characteristic of weak-lensing distributions~\cite{Schneider1999-tk}. The number of halos contributing within the wave-optics volume is
\bea\label{eq:Mcut}
    &\langle N_{\rm halos}\rangle_{M>M_{\rm cut}}\\
    &=\int_0^{z_s}\td z\int_{M_{\rm cut}}^{M_{\rm v,max}}\td M\, \,\frac{\pi (N\,r_f(f_{\rm min}))^2(1+z)^2}{H(z)}\frac{\td n}{\td M}\, ,
\eea
where we take the wave-optics radius to be $N$ times the Fresnel scale. This choice inevitably includes halos in the geometric limit, but avoids spurious finite simulation-volume problems. $N$ is chosen large enough that the results are insensitive to the cut; for simulations we take $N=10$. The contribution from a single halo to $\Delta F$, depending on the impact parameter $r_c$ from the line of sight, is
\bea\label{eq:DF1}
    \Delta F_1(w;M,r_c)
    &= \frac{iw}{\pi}\int_0^{\infty}\!\frac{\td k}{k}\,\hat\kappa_{\rm NFW}(k;M,z)\,\\
    &\times J_0(k\,r_c)\,
    \Bigl(e^{-ik^2/2w}-1\Bigr)\,,
\eea
and the mean field of the sampled population is removed once per realisation,
\be\label{eq:DF1mean}
    \Delta F(w) \;=\; \sum_i \Delta F_1(w;M_i,r_{c,i})
    \;-\; \langle N_{\rm halos}\rangle\,\overline{\Delta F_1}(w;R)\,,
\ee
where $R \equiv N$ is the dimensionless radius of the wave-optics disk, $\langle N_{\rm halos}\rangle$ is given by Eq.~\eqref{eq:Mcut}, and $\overline{\Delta F_1}$ is the average of Eq.~\eqref{eq:DF1} over impact parameters in the disk and over the sampled mass function, obtained by $J_0(k\,r_c)\to 2J_1(kR)/(kR)$. This deterministic subtraction embeds the finite disk in the ensemble-mean background: it enforces $\langle F\rangle=1$ while leaving the Poisson variance $\bar n\,|\hat\kappa_{\rm NFW}|^2$ of Eq.~\eqref{eq:cov_def} untouched, whereas a per-halo subtraction would scale with the Poisson-random $N_{\rm halos}$ and bias the small-$k$ variance.\footnote{For $R$ much larger than the halo extent the subtraction reduces to $\hat\kappa(z)\,\Delta F_R(w)$, the response of a uniform disk of convergence $\hat\kappa$ [Eq.~\eqref{eq:kappahat}], with the closed form $\Delta F_R = 1 - e^{iT} - iT\,E_1(-iT)$, $T=wR^2/2$, so that $1-\Delta F_R(w)=\mathcal{O}\!\bigl(2/wR^2\bigr)$: a real, nearly frequency-independent mean, coinciding with the geometric ($n=0$) mode that is projected out of the detection statistic, which reduces to the cosmic-mean ($k\to0$) subtraction as $R\to\infty$ and therefore does not affect the forecast.}

The Gaussian limit is accurate only in the mass bins for which many halos populate the Fresnel volume; the remaining, heavy-tail bins must be sampled halo-by-halo. We impose a single mass threshold $M_{\rm cut}$ that splits the mass integral into a Gaussian piece and an explicit Poisson piece,
\bea\label{eq:DFsplit}
    \Delta F(w) &=\; \Delta F_{\rm G}(w;\,M_{\rm v,min},M_{\rm cut}) \\
    +\; &\Delta F_{\rm P}(w;\,M_{\rm cut},M_{\rm v,max})\,.
\eea
The Gaussian contribution $\Delta F_{\rm G}$ is drawn via a Cholesky factorisation of the covariance from Eq.~\eqref{eq:cov_def} evaluated on the mass range $[M_{\rm v,min},M_{\rm cut}]$. The Poisson contribution $\Delta F_{\rm P}$ is the sum $\sum_i \Delta F_1(w;M_i,r_{c,i})$ over an explicit halo catalogue drawn by Poisson-sampling the mass function on $[M_{\rm cut},M_{\rm v,max}]$ and the Fresnel volume, with the deterministic mean-field subtraction of Eq.~\eqref{eq:DF1mean} restricted to the same mass range. For the numerical implementation, we use $N=10$ and the Gaussian approximation for masses that satisfy $\langle N_{\rm halos}\rangle\gtrsim100$.

\newpage
\section{Effects of the halo concentration}\label{sec:concentration}
Wave-optics lensing is very sensitive to the morphology of the halos, with more concentrated objects (i.e. higher central density) producing stronger diffractive features~\cite{Savastano:2023spl,Brando:2024inp}. Each halo enters the convergence power spectrum, Eq.~\eqref{eq:Pkappa_1h_z}, only through the analytic NFW Hankel transform $\hat\kappa_{\rm NFW}(q;\kappa_0,b)$ of Eq.~\eqref{eq:kappa_hat_NFW}, which is fixed by two dimensionless numbers: the convergence normalisation $\kappa_0=\rho_s r_s/\Sigma_{\rm cr}$ and the inverse scale radius $b=\xi_0/r_s=r_f(f_{\rm min})/r_s$. Both inherit their dependence on the halo's internal structure from a single quantity, the concentration $c\equiv r_{200}/r_s$. Here we make that dependence explicit. For an NFW profile of virial mass $M$ (defined at $200\,\rho_c$), the virial radius
\be\label{eq:r200}
    r_{\rm v}(M,z)=\left[\frac{3M}{4\pi\,\Delta\,\rho_c(z)}\right]^{1/3}
\ee
$(\Delta=200)$ is fixed by mass and redshift alone, whereas the scale radius and characteristic density carry all of the concentration dependence,
\be\label{eq:rs_rhos}
    r_s=\frac{r_{200}}{c}\,,\qquad
    \rho_s=\frac{200}{3}\,\rho_c(z)\,\frac{c^3}{\Pi(c)}\,,\quad
    \Pi(c)\equiv\ln(1+c)-\frac{c}{1+c}\,.
\ee
Substituting Eq.~\eqref{eq:rs_rhos} into the two lensing parameters at fixed $(M,z)$ gives their concentration scaling,
\begin{align}
    b&=\frac{\xi_0}{r_s}=\frac{\xi_0}{r_{200}}\,c\;\;\propto\;c\,,\label{eq:b_of_c}\\
    \kappa_0&=\frac{\rho_s r_s}{\Sigma_{\rm cr}}
    =\frac{200}{3}\,\frac{\rho_c\,r_{200}}{\Sigma_{\rm cr}}\,\frac{c^2}{\Pi(c)}
    \;\;\propto\;\frac{c^2}{\Pi(c)}\,.\label{eq:kappa0_of_c}
\end{align}
A more concentrated halo therefore (i) has a smaller scale radius, so its lensing response is pushed to larger dimensionless wavenumbers $q\sim b$ (closer to the geometric-optics regime at fixed frequency), and (ii) has a deeper central convergence, raising $\kappa_0$. The overall amplitude of the transform scales as $\kappa_0/b^2\propto1/\Pi(c)$ up to a logarithm, so it grows only logarithmically with $c$; but the per-halo contribution to the variance, $\propto\kappa_0^2$, grows as $c^4/\Pi(c)^2$, so the concentration is a leading-order control on the strength of the signal at fixed halo abundance.

The concentration is not a free parameter: it is set by the amplitude of fluctuations on the halo mass scale through the peak height $\nu(M,z)=\delta_c(z)/\sigma(M)$. We adopt the Okoli \& Afshordi relation~\citep{2016MNRAS.456.3068O}, since being analytical can be extrapolated in redshift and for different DM models 
\be\label{eq:cOA}
    \log_{10}c(M,z)=1.09+0.78\,\ln y\,,\qquad
    y=\Bigl[0.42+0.2\,\nu^{-1.23}\Bigr]\bigl[H(z)\,t(z)\bigr]^{-2/3}\,,
\ee
with $\nu=\delta_c(z)/\sigma(M)$ and $t(z)$ the look-back time. The key point for the beyond-CDM analysis is that $\sigma(M)$ is computed from the \emph{modified} power spectrum $\mathcal{P}(k;k_0,k_{\rm cut})$ of Eq.~\eqref{eq:Pmod_supp}, so the concentration is adjusted self-consistently with the halo abundance, rather than being imported from a CDM calibration.

There are two caveats with this approach. In the suppression models (Fuzzy/Warm DM, $k_{\rm cut}=k_0$), the small-scale variance $\sigma(M)$ drops towards the cut-off mass, so the few surviving low-mass halos are also \emph{less} concentrated. Through Eqs.~\eqref{eq:b_of_c}--\eqref{eq:kappa0_of_c}, this lowers both $\kappa_0$ and $b$, weakening their lensing imprint beyond the reduction in number alone. Conversely, the white-noise plateau of the enhancement models (axion miniclusters, PBHs, $k_{\rm cut}>k_0$) raises $\sigma(M)$ at small $M$, lowering $\nu$ and increasing $c$, so the additional halos are also more concentrated and contribute more variance per object. This concentration response amplifies the abundance response in the direction that helps detection for enhancements, which is why the constraints quoted in the main text are conservative. For Fuzzy DM, the effect is even stronger in that direction: the wave nature of the field produces a central soliton core~\cite{Schive:2014dra} that lowers the inner density below the NFW value assumed in Eq.~\eqref{eq:rs_rhos}, so our benchmarks should be read as lower bounds on the achievable signal. Of course, for the primordial black hole case, the point mass nature of the lens will further enhance detectability; we leave this more detailed analysis for future work.

\begin{figure}[t]
    \centering
    \includegraphics[width=0.98\textwidth]{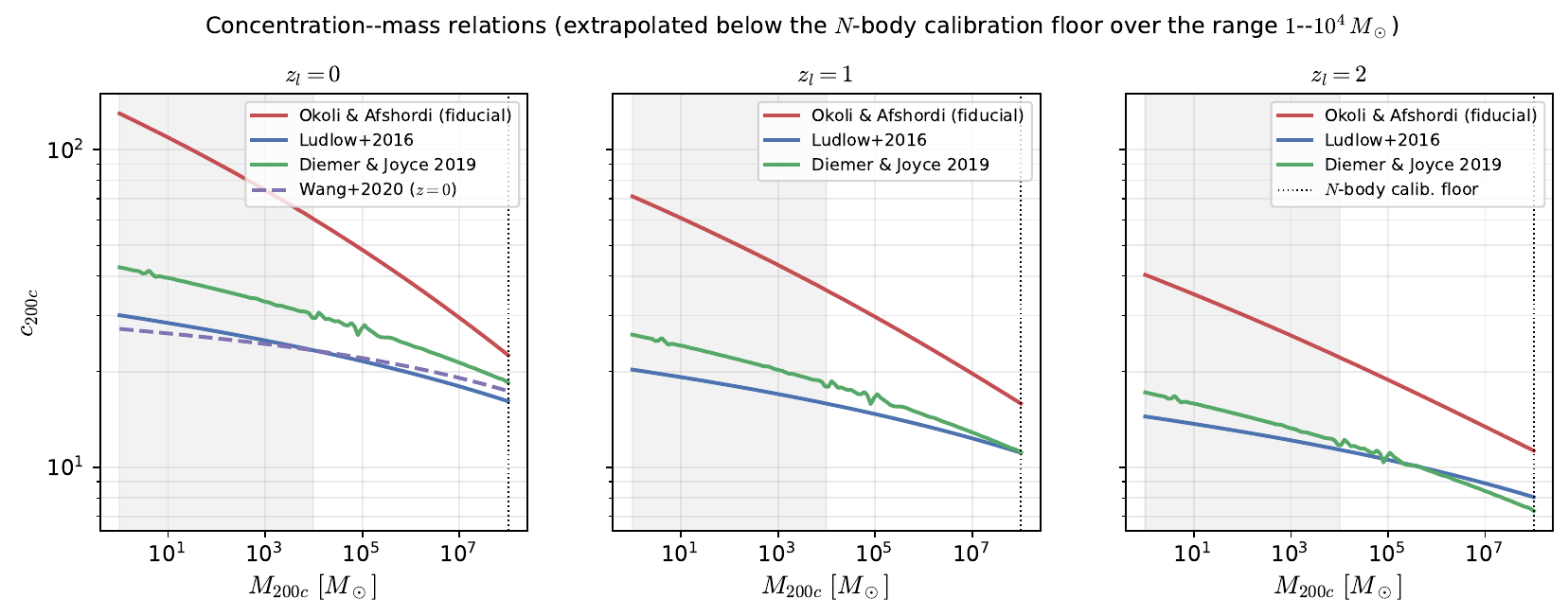}
    \caption{Concentration--mass relations $c_{200c}(M)$ at lens redshifts $z_l=0,1,2$ for the prescriptions considered: the fiducial Okoli \& Afshordi~\citep{2016MNRAS.456.3068O}, Ludlow et al.\ 2016~\citep{Ludlow:2016ifl}, Diemer \& Joyce 2019~\citep{Diemer:2018vmz}, and (at $z=0$ only) Wang et al.\ 2020~\citep{Wang:2019ftp}. The shaded band marks the LISA-relevant halo range $1$--$10^4\,\Msun$ and the dotted line a representative $N$-body calibration floor: all relations are extrapolations over the LISA range. Okoli \& Afshordi sits systematically highest (most concentrated, most optimistic).}
    \label{fig:conc_relations}
\end{figure}

The relevant LISA halos ($1$--$10^4\,\Msun$) lie one to several decades below the smallest masses resolved in the $N$-body simulations that calibrate every published concentration--mass relation~\citep{Bullock:1999he,Wechsler:2001cs,Prada:2011jf,Diemer:2014gba,Correa:2015dva}, so $c(M,z)$ over this range is a genuine extrapolation and the dominant astrophysical uncertainty on the signal strength. To bracket it, we recompute the full Dark-Timbre detectability of the fiducial test binary ($\Mc=10^6\,\Msun$, $\eta=0.25$, $z_s=2$, $5$~yr; the same event as Sec.~\ref{sec:detstat}) under four alternative prescriptions, varying \emph{only} the concentration entering Eqs.~\eqref{eq:b_of_c}--\eqref{eq:kappa0_of_c} and holding the mass function, $\sigma(M)$, the frequency band, the noise-weighted inner product and the five-parameter waveform manifold fixed. The relations are shown in Fig.~\ref{fig:conc_relations} and the resulting evidence in Table~\ref{tab:concentration}.

\begin{table}[t]
\centering
\caption{Concentration dependence of the Dark-Timbre detectability for the fiducial test binary ($\Mc=10^6\,\Msun$, $\eta=0.25$, $z_s=2$, $5$~yr), changing only the concentration--mass relation. $c(10\,\Msun,z{=}1)$ is a representative concentration in the LISA halo range and $f_{\rm surv}=\sum_n s_n/\sum_n\sigma_n^2$ the fraction of the lensing power that survives the projection over the binary parameters (Sec.~\ref{sec:detstat}); the per-event significance $\rho$ and catalogue size $N_{\rm bin}=(S/\rho)^2$ are quoted relative to the fiducial Okoli \& Afshordi (OA) value of Sec.~\ref{sec:detstat}. Wang+20 is $z=0$-calibrated and listed as an extrapolation.}
\label{tab:concentration}
\begin{tabular}{|l|c|c|c|c|}
\hline
Relation & $c\,(10\,\Msun,z{=}1)$ & $f_{\rm surv}$ & $\rho/\rho_{\rm OA}$ & $N_{\rm bin}/N_{\rm bin}^{\rm OA}$\\
\hline
Okoli \& Afshordi (fiducial) & $60.9$ & $5.8\times10^{-4}$ & $1.00$ & $1.0$\\
Wang+20 ($z=0$, extrap.)     & $26.4$ & $2.9\times10^{-4}$ & $0.46$ & $4.7$\\
Diemer \& Joyce 2019         & $24.1$ & $2.9\times10^{-4}$ & $0.33$ & $9.3$\\
Ludlow+2016                  & $19.2$ & $2.3\times10^{-4}$ & $0.24$ & $16.8$\\
\hline
\end{tabular}
\end{table}

The trend follows directly from the scaling of Eq.~\eqref{eq:kappa0_of_c}: a lower concentration lowers $\kappa_0\propto c^2/\Pi(c)$, hence the per-halo variance $\propto\kappa_0^2\propto c^4/\Pi(c)^2$ and the eigenvalues $\sigma_n^2$ of the lensing covariance. Okoli \& Afshordi is the most concentrated, and hence the most optimistic, of the relations considered, giving the largest per-event significance and the smallest $N_{\rm bin}$: the less-optimistic CDM-calibrated fits require roughly $9$ (Diemer \& Joyce) to $17$ (Ludlow) times more binaries to reach the same evidence. The variance proxy $c^4/\mu(c)^2$ accounts for the ranking quantitatively, dropping to $\sim\!4$--$10\%$ of the  Okoli \& Afshordi value for the CDM fits over the LISA halo range.

This spread does not weaken the constraints quoted in the main text: because Okoli \& Afshordi is the optimistic end, the CDM forecasts should be read as upper bounds on the achievable evidence from the concentration side. On the other hand,  for the enhancements on the matter power spectrum that we have considered, our results should be read as a lower bound since these models predict more concentrated halos, i.e. point-like in the PBH case or with a solitonic core in the axion minicluster case. We nonetheless retain Okoli \& Afshordi as the fiducial throughout because it is \emph{analytic}: it can be extrapolated in redshift and, crucially, recomputed self-consistently from the modified power spectrum $\sigma(M)$ of Eq.~\eqref{eq:Pmod_supp} for each beyond-CDM model, which the simulation-calibrated fits cannot.

\newpage
\section{Detection statistic and catalogue combination}\label{sec:detstat}
\begin{figure}[h!]
    \centering
    \includegraphics[width=0.9\textwidth]{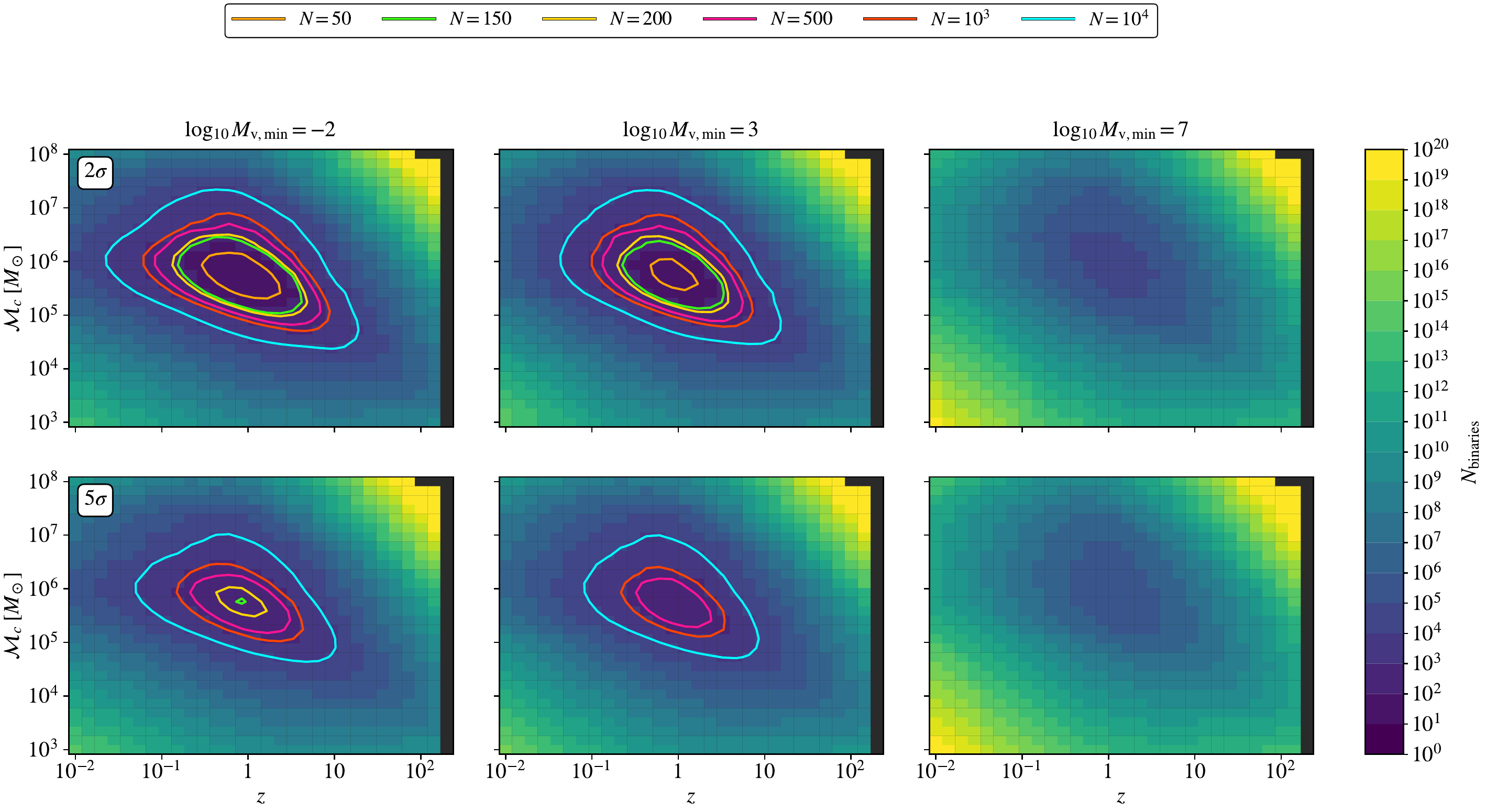}
    \caption{Number of binaries required, assuming one year of observation and equal-mass components for LISA, as a function of the chirp mass and the binary's redshift for $2\sigma$ C.L. top panels and $5\sigma$ for the lower ones. In the first column, we can see the necessary number of binaries for the CDM halo mass function down to $10^{-2}\,M_{\odot}$, in the second column down to $10^3\,M_{\odot}$ and in the last column cutting the halos at $10^7\,M_{\odot}$.}
    \label{fig:N_bin}
\end{figure}

The Dark Timbre is a zero-mean stochastic signal, $\left(\langle\Delta F\rangle=0\right)$. There is therefore no unique waveform to match-filter against: the lensing imprint cannot be detected through its mean, only through the \emph{excess variance}. Here we construct the corresponding likelihood-ratio statistic, derive its expected significance, and show how it is combined over a catalogue of sources. We started by decomposing the amplification factor in the basis
\be
\Delta F(f)=\sum_n c_n\psi_n(f)\, ,
\ee
with $c_n\sim\mathcal{CN}(0,\sigma_n^2)$ quoted in Eq.~\eqref{eq:KL}. The complex functions that make the basis $\psi_n(f)$ come from fulfilling that are eigenfunctions to the following operator
\begin{equation}
    \hat{\mathbb{C}}\,\psi_n
    \;=\; \sigma_n^{2}\,\psi_n,
    \qquad
    \langle \psi_m, \psi_n\rangle_w
    \;=\; \delta_{mn}.
\label{eq:eigen_problem}
\end{equation}
with the operator $\hat{\mathbb{C}}$ defined as
\begin{equation}
    \big(\hat{\mathbb{C}}\,\psi\big)(f)
    \;\equiv\;
    4\int df'\; W(f')\, C(f,f')\,\psi(f'),
\label{eq:Chat_def}
\end{equation}
whose factor of $4$ matches the normalisation of $\langle\cdot,\cdot\rangle_w$ from Eq.\eqref{eq:inner}. We say that each equation is \textit{statistically} independent because, in virtue of Mercer's theorem, the covariance, which contains all the information about the stochastic diffraction in the Gaussian limit, admits the expansion
\begin{equation}
    C(f,f')
    \;=\; \sum_n \sigma_n^{2}\,\psi_n(f)\,\psi_n^{*}(f'),
\label{eq:mercer}
\end{equation}
Now we will discuss how this translates into an excess of fluctuations in the shape of the eigenmodes $\psi_n$. Working in the Karhunen--Lo\`eve basis of Eq.~\eqref{eq:KL}, let us consider a single event with frequency-domain data
\be\label{eq:data_model}
    \tilde d(f) \;=\; \tilde h(f,\bm\theta)\,\bigl[1+\Delta F(f)\bigr]
    \;+\; \tilde n(f)\, ,
\ee
where $\tilde h(f,\bm\theta)$ is the unlensed waveform and $\tilde n(f)$ the
detector noise, stationary and Gaussian with
$\langle\tilde n(f)\,\tilde n^{*}(f')\rangle=\tfrac12 S_n(f)\,\delta(f-f')$.
The residual, after subtraction of the maximum-likelihood unlensed template
$\tilde h(\bm\theta_0)$, projects onto a set of statistically independent mode
amplitudes,
\be\label{eq:dn_def}
    d_n \;\equiv\; \mathrm{Re}\,\Bigl\langle \psi_n,\;
    \frac{\tilde d-\tilde h(\bm\theta_0)}{\tilde h}\Bigr\rangle_W\, ,
\ee
the data-side counterpart of the theory coefficients $c_n$ of
Eq.~\eqref{eq:KL}. No explicit noise rescaling is needed,  each real quadrature carries unit
noise variance, so in the absence of lensing $d_n\sim\mathcal{N}(0,1)$.
The two competing hypotheses then differ only in the variance of that quadrature,
\be\label{eq:hypotheses}
    \mathcal{H}_0:\ d_n\sim\mathcal{N}(0,1)\, ,
    \qquad
    \mathcal{H}_1:\ d_n\sim\mathcal{N}(0,1+s_n)\, .
\ee
Since $\mathcal{H}_0$ and $\mathcal{H}_1$ are zero-mean Gaussians that differ only in their covariance, the log-likelihood ratio reduces to a sum of independent per-mode excess-power tests,
\be\label{eq:logLR}
    2\ln\frac{\mathcal{L}_1}{\mathcal{L}_0}
    \;=\; \sum_n\Bigl[\frac{s_n}{1+s_n}\,d_n^{2}-\ln(1+s_n)\Bigr]\, .
\ee
The $s_n$ can be computed, as discussed in the main text, as $s_n=(1-\Lambda_n)\sigma^2_n$. The expected separation between the hypotheses is set by the Kullback--Leibler divergence per mode, i.e.\ half the expectation of Eq.~\eqref{eq:logLR} under $\mathcal{H}_1$. For two real Gaussians of equal mean and variances $1$ and $1+s_n$,
\be\label{eq:KLdiv}
    D_n \;=\; \tfrac12\Bigl[s_n-\ln(1+s_n)\Bigr]
    \;\simeq\; \tfrac14\,s_n^{2}\, .
\ee
The expected detection significance squared is the total divergence,
\be\label{eq:rho_det}
    \rho^{2} \;=\; \sum_n D_n
    \;\simeq\; \frac14\sum_n s_n^{2}
    \;=\; \frac14\sum_n \sigma_n^{4}\,(1-\Lambda_n)^{2}\, .
\ee
The quartic dependence on $\sigma_n$ is the hallmark of a second-moment (variance) detection, in contrast with the linear scaling of the total signal power $\langle\Delta\chi^2\rangle=\sum_n s_n$ of Eq.~\eqref{eq:chi2}. Figure~\ref{fig:spectral_decomp} shows that, after marginalisation, the detectable power is concentrated in the first few modes. We evaluate Eq.~\eqref{eq:rho_det} directly for every event. If a single mode dominated the sum, $\sum_n s_n^{2}\simeq\bigl(\sum_n s_n\bigr)^{2}=\langle\Delta\chi^2\rangle^{2}$ and the per-event significance would reduce to the compact form
\be\label{eq:rho_chi2}
    \rho \;\simeq\; \tfrac12\,\langle\Delta\chi^2\rangle\, ;
\ee
in practice, however, the leading modes carry comparable power, $\bigl(\sum_n s_n\bigr)^{2}/\sum_n s_n^{2}\simeq3$ for the fiducial event below, so Eq.~\eqref{eq:rho_chi2} would overestimate $\rho$ by a factor $\simeq\sqrt{3}$ and we do not use it. For the fiducial LISA binary ($\mathcal{M}_c=10^6\,M_\odot$, $z_s=2$, $5$ years before merger) we find $\langle\Delta\chi^2\rangle=\sum_n s_n=0.89$ and $\sum_n s_n^{2}=0.27$, hence $\rho\simeq0.26$ per event. The Dark Timbre is present in every source and the per-event statistics are independent, so significances add in quadrature,
\be\label{eq:sigmaEvi_supp}
    \sigma^2_{\rm Evi.}\;=\;\sum_i\rho_i^{2}
    \;=\;\frac14\sum_i\sum_n s_{n,i}^{2}\, .
\ee
A detection at significance $S$ then requires a catalogue of comparable events of size
\be\label{eq:Nbin_supp}
    N_{\rm bin.}\;\simeq\;\Bigl(\frac{S}{\rho}\Bigr)^{2}
    \;=\;\frac{4S^{2}}{\sum_n s_n^{2}}\, .
\ee
For $S=2$ this yields $N_{\rm bin.}\simeq60$ for the fiducial event. Because the combination is quadratic in the per-mode powers $s_{n,i}$, the total significance is dominated by the loudest events in a catalogue rather than by the raw number of detections. In Fig.~\ref{fig:N_bin}, we show how many binaries are necessary to detect at the $2\sigma$ C.L. the Dark Timbre considering different cuts of the halo mass function. We can see how for $M_{\rm v\,,min}=10^{-2}M_\odot$ fewer binaries are required, $M_{\rm v\,,min}=10^{3}M_\odot$ increases the required number of detections, while $M_{\rm v\,,min}=10^{7}M_\odot$ demands a huge number of sources. 

\begin{figure}[h!]
    \centering
     \includegraphics[width=0.7\textwidth]{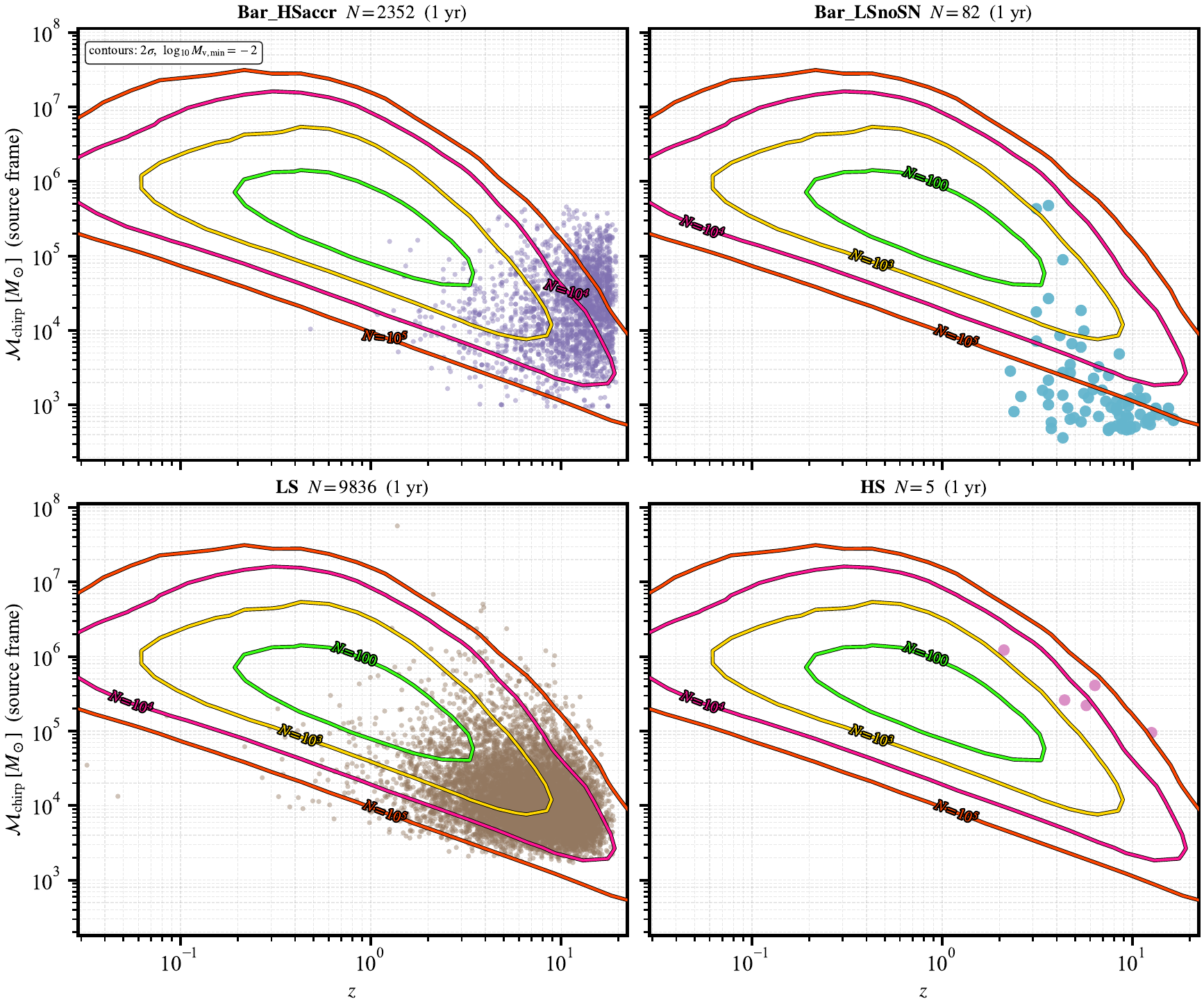}
    \caption{Same contours as in Fig.~\ref{fig:N_bin} mergers but overlaying all the mergers in the merger rates we are considering. Notice that not all the merger have equal masses, we project them for visualisation purposes.}
    \label{fig:N_bin_2}
\end{figure}

 This is in accordance with the conclusions that the Dark Timbre signal is dominated by halos of $(10-10^4)\, M_{\odot}$, and receives negligible contribution from halos with $M_{\rm v}<1\,M_{\odot}$. We can see how $50$ of the loudest binaries are needed to detect the Dark Timbre at $2\sigma$ while around $500$ are needed to detect it at $5\sigma$. In Fig.~\ref{fig:N_bin_2} we compare the per-event significance to the expected properties of events under different black hole formation scenarios, assuming 1 year of observation. Notice that the comparison is only indicative since the contours assume equal mass, which is not true for all the binaries we display. The mass ratio is accounted for when we evaluate the significance of synthetic source populations.

\emph{Detection in practice---}The statistic above admits an equivalent parameter-estimation (PE) formulation: treating the Dark-Timbre coefficients $c_n$ as nuisance parameters with the theory-predicted prior $c_n\sim\mathcal{CN}(0,\sigma_n^2)$ and marginalising over them adds the lensing covariance to the data, so that the Bayesian evidence of $\mathcal{H}_1$ against $\mathcal{H}_0$ [Eq.~\eqref{eq:hypotheses}] reproduces the likelihood ratio of Eq.~\eqref{eq:logLR} (cf.\ the PE-based treatment of~\cite{Zumalacarregui:2026uqs}). The discriminating power lies entirely in the predicted variances $\sigma_n^2$: the basis is generic and could absorb any waveform deviation, but it is the CDM (or modified-DM) prior on the $c_n$ that turns the projection into a test of a specific model, rather than a fit of free coefficients. Because a single event provides only one realisation of the field, the variance is sampled across the catalogue, each source contributes one measurement of the whitened residual power in the surviving modes, and the significances combine as in Eq.~\eqref{eq:sigmaEvi_supp}, in the spirit of stochastic-signal searches such as those in pulsar-timing arrays~\cite{NANOGrav:2023gor,EPTA:2023sfo} that also have vanishing mean. A dedicated data-analysis pipeline for this excess-variance observable in resolved sources is left to future work.

\newpage

\section{Beyond Cold Dark Matter}\label{sec:beyondCDM}

In the main text, we parametrised departures from the CDM matter power spectrum with the two-scale step
\bea\label{eq:Pmod_supp}
    \mathcal{P}(k;\,k_0,k_{\rm cut}) &= \mathcal{P}_{\rm CDM}(k)\,\Theta(k_0-k)\\
    &+ \mathcal{P}_{\rm CDM}(k_0)\,\Theta(k_{\rm cut}-k)\,\Theta(k-k_0)\, ,
\eea
with $k_{\rm cut}\geq k_0$. The parameterization describes a drop in the power spectrum at $k_{\rm cut}$, with the possibility of enhancing the power spectrum through a plateau at $k_0<k_{\rm cut}$. The CDM case is recovered when $k_{\rm cut} = k_0 \to \infty$. This single shape captures the two qualitatively different ways in which a DM model can modify small-scale structure. When $k_{\rm cut}=k_0$ the plateau collapses and Eq.~\eqref{eq:Pmod_supp} reduces to CDM with a sharp UV cut-off at $k_0$: this is the \emph{suppression} regime, appropriate for models with a small-scale free-streaming or quantum-pressure cut-off such as Fuzzy or Warm DM. When $k_{\rm cut}>k_0$ the spectrum acquires a flat, white-noise plateau $\mathcal{P}_{\rm CDM}(k_0)$ between $k_0$ and $k_{\rm cut}$: this is the \emph{enhancement} regime, appropriate for models that inject extra small-scale power, such as axion miniclusters or primordial black holes (PBHs). The purpose of this section is to map the physical parameters of each model onto $(k_0,k_{\rm cut})$, so that the constraints of Table~\ref{tab:results} can be read directly in terms of particle masses and abundances, and to compare these analytic maps with the numerical implementation used to produce Fig.~\ref{fig:ko:cut}.

\subsection{Suppressed small-scale structure: Fuzzy Dark Matter}

In Fuzzy Dark Matter (FDM) the DM is an ultralight bosonic field of mass $m_{\rm FDM}=\mathcal{O}(10^{-20}\,{\rm eV})$ whose de Broglie wavelength is of astrophysical size~\cite{Hui:2016ltb,Hui:2021tkt,Ferreira:2020fam,Singh:2025uvp}. The modification to the sound speed, $c_s\sim k/(4a^2m_{\rm FDM}^2)$, generates a non-zero Jeans scale below which perturbations cannot grow~\cite{Hu:2000ke,Marsh:2015xka},
\be\label{eq:kJ_FDM}
    k_{\rm J} = \frac{66.5\,{\rm Mpc}^{-1}}{(1+z_{\rm eq})^{1/4}}
    \left[\frac{\Omega_{\rm FDM}h^2}{0.12}\right]^{\frac14}
    \left[\frac{m_{\rm FDM}}{10^{-22}\,{\rm eV}}\right]^{\frac12}\, .
\ee
Power at $k\gtrsim k_{\rm J}$ is damped, with a transfer function accurately fit by~\cite{Passaglia:2022bcr}, $T_{\rm F}=\sin x^{n}/[x^{n}(1+Bx^{6-n})]$ and $x=A\,k/k_{\rm J}$, where $A=2.22\,m_{22}^{1/25+\ln m_{22}/1000}$, $B=0.16\,m_{22}^{-1/20}$ and $m_{22}\equiv m_{\rm FDM}/10^{-22}\,{\rm eV}$.
Because the FDM spectrum is simply cut, we take the suppression branch $k_{\rm cut}=k_0$ of Eq.~\eqref{eq:Pmod_supp} and identify the step scale with the Jeans scale,
\be\label{eq:k0_FDM}
    k_0^{\rm FDM}\;=\;k_{\rm cut}^{\rm FDM}\;\simeq\;k_{\rm J}(m_{\rm FDM})\, .
\ee
The analogous suppression map for thermal Warm DM follows identically from its free-streaming transfer function~\citep{Bode:2000gq,Viel:2005qj}, with $k_0$ set by the half-mode of $T_{\rm W}=[1+(\alpha k)^{2\mu}]^{-5/\mu}$.

\subsection{Enhanced small-scale structure: axion miniclusters and PBHs}

If the Peccei--Quinn $U(1)$ symmetry breaks \emph{after} inflation, the axion field takes uncorrelated values in causally disconnected patches. The resulting $\mathcal{O}(1)$ density fluctuations at the horizon scale when the axion begins to oscillate seed gravitationally bound \emph{axion miniclusters} via the Kibble mechanism~\citep{Kolb:1993zz,Kolb:1993hw,Vaquero:2018tib,Eggemeier:2019khm}. Being set by a single causal patch, these fluctuations are uncorrelated and contribute a scale-invariant white-noise term to the matter power spectrum~\citep{Fairbairn:2017sil,Feix:2019lpo,Ellis:2022grh}, exactly the plateau modelled by Eq.~\eqref{eq:Pmod_supp}. The onset and cut-off scales are
\begin{align}
    k_0^{\rm AMC}&\equiv k_c\;\approx\;3\,h\,{\rm Mpc}^{-1}\left[\frac{m_a}{10^{-18}\,{\rm eV}}\right]^{0.6}\, ,\label{eq:k0_AMC}\\
    k_{\rm cut}^{\rm AMC}&\;\simeq\;300\,{\rm Mpc}^{-1}\sqrt{\frac{m_a}{10^{-18}\,{\rm eV}}}\, ,\label{eq:kcut_AMC}
\end{align}
where $m_a$ is the axion mass and $k_{\rm cut}$ is the comoving scale that re-entered the horizon at the onset of oscillations. In the relevant parameter space $k_0\ll k_{\rm cut}$, so the plateau is wide and the enhancement branch $k_{\rm cut}>k_0$ applies; the Jeans scale lies beyond the cut, $k_{\rm J}>k_{\rm cut}$.

A subdominant population of PBHs~\citep{Carr:2020xqk,Green:2020jor} produces an analogous white-noise enhancement: discreteness (Poisson) fluctuations in the PBH number density add isocurvature power on small scales~\citep{Afshordi:2003zb,Inman:2019wvr,DeLuca:2020jug}, with onset
\be\label{eq:k0_PBH}
    k_0^{\rm PBH}\equiv k_c\;\approx\;6\,h\,{\rm Mpc}^{-1}\left[\frac{f_{\rm PBH}\,m_{\rm PBH}}{10^4\,\Msun}\right]^{-0.4}\, ,
\ee
where $f_{\rm PBH}$ is the PBH fraction of the DM and $m_{\rm PBH}$ its mass. The plateau is truncated at the mean PBH separation~\cite{Hutsi:2022fzw}, below which there is roughly one PBH per comoving volume and the individual-seed effect~\citep{Carr:2018rid,Cappelluti:2021usg} takes over,
\be\label{eq:kcut_PBH}
    k_{\rm cut}^{\rm PBH}\;=\;900\,h\,{\rm Mpc}^{-1}\left[f_{\rm PBH}\,\frac{10^4\,\Msun}{m_{\rm PBH}}\right]^{1/3}\, .
\ee
The hierarchy $k_0<k_{\rm cut}$ holds for $f_{\rm PBH}>1.1\times10^{-3}\,(m_{\rm PBH}/10^4\,\Msun)^{-0.09}$. 
\begin{figure*}
    \centering
    \includegraphics[width=\textwidth]{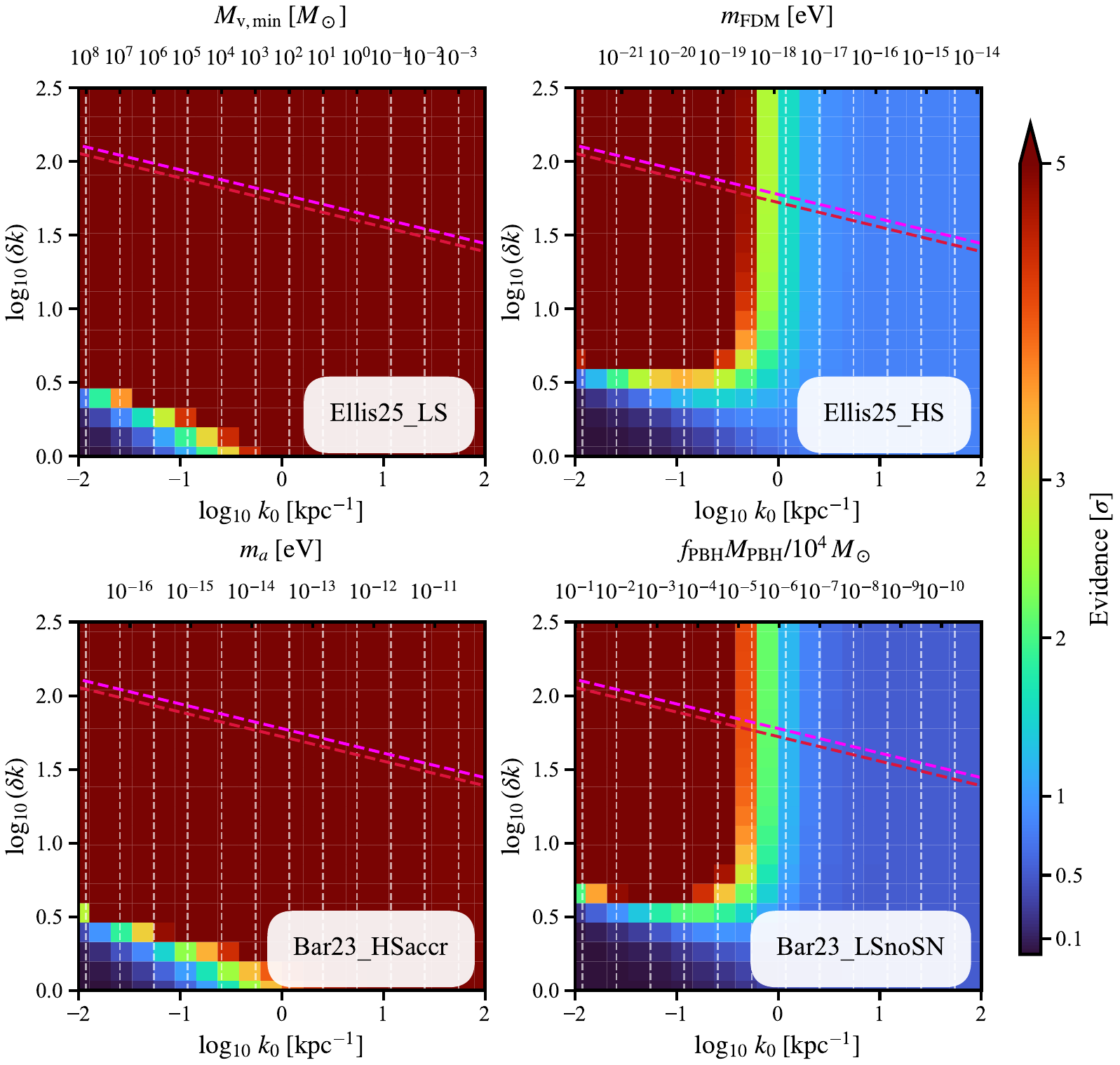}
    \caption{Expected evidence $[\sigma]$ for a modified matter power spectrum, Eq.~\eqref{eq:Pmod_supp}, after $10$ years of LISA observation, as a function of the onset scale $\log_{10}k_0$ and the plateau width $\delta k\equiv k_{\rm cut}/k_0$ (plotted as $\log_{10}\delta k$, matching the vertical axis), for four merger-rate models. The bottom edge $\log_{10}\delta k=0$ is the suppression (Fuzzy/Warm DM) limit, while $\log_{10}\delta k>0$ corresponds to the enhancement (axion-minicluster/PBH) regime. The dashed lines mark the trajectories of axion miniclusters [crimson, lower; Eqs.~\eqref{eq:k0_AMC}--\eqref{eq:kcut_AMC}] and PBHs [magenta, upper; Eqs.~\eqref{eq:k0_PBH}--\eqref{eq:kcut_PBH}] in this plane. The top axis of each panel translates $k_0$ into the corresponding physical parameter through Eqs.~\eqref{eq:k0_FDM}, \eqref{eq:k0_AMC} and \eqref{eq:k0_PBH}.}
    \label{fig:ko:cut}
\end{figure*}

Fig.~\ref{fig:ko:cut} shows a more detailed constrained plot. We can see that for the $4$ different merger rates we have considered, the evidence after $10$ years that can be achieved for different values of $k_0\, ,k_{\rm cut}$ expressed as $\log_{10}(\delta k)\equiv\log_{10}(k_{\rm cut}/k_0)$  for maximising the space in each of the plots. As expected, both AMCs and PBHs are lines in this plane: the crimson (lower) dashed line is the AMC trajectory, with the axion mass shown as a reference in the lower left panel, and the magenta (upper) dashed line is the PBH trajectory, with the total density mass shown in the lower right panel. We can see that both lines are very similar in terms of $\delta k(k_0)$ even if they map to very different physical phenomena. In the top-left panel, we show the halo masses at the scale $k_0$, through $M_{\rm v}=\tfrac{4\pi}{3}\bar\rho_{\rm m,0}\,k_0^{-3}$, which can be mapped to a hard-UV cut by selecting halos below that mass. As we have discussed, this can be approximately realised in FDM scenarios, for which we show the mass scales in the top right panel. The two panels to the left show that it is possible to measure a low-mass cut-off of the halo mass function along the suppression line $\log_{10}\delta k=0$; standard CDM is recovered in the limit $k_0\to\infty$ of that line. In general, the evidence alongside the $\log_{10}\delta k=0$ line is equivalent to Fig.~\ref{fig:CDM_post}, where the $x$-axis is expressed as $M_{\rm v,min}$ shown with the white dashed lines. We can see that the top left corner of all the plots, corresponding to enhancements at $(10^8-10^4)\,M_{\odot}$, are always detectable: the white-noise plateau raises $\sigma(M)$ in the mass range where LISA is most sensitive, boosting both the abundance of these halos and, self-consistently, their concentration (Sec.~\ref{sec:concentration}), which amplifies the per-halo signal. For lower halo masses, corresponding to $(10^2-1)\,M_{\odot}$, whether this is detectable or not depends strongly on the luminosity of the different merger rates. Finally, let us notice, which is more evident by looking at Fig.~\ref{fig:CDM_post}, that for the lowest halo masses $<10\,M_{\odot}$ we are no longer sensitive to these masses, so the evidence just flattens; that is why we report the constraints up to these scales. 

\section{Spin precession $\&$ higher modes}\label{sec:spin}

\begin{figure}[t]
    \centering
    \includegraphics[width=0.60\textwidth]{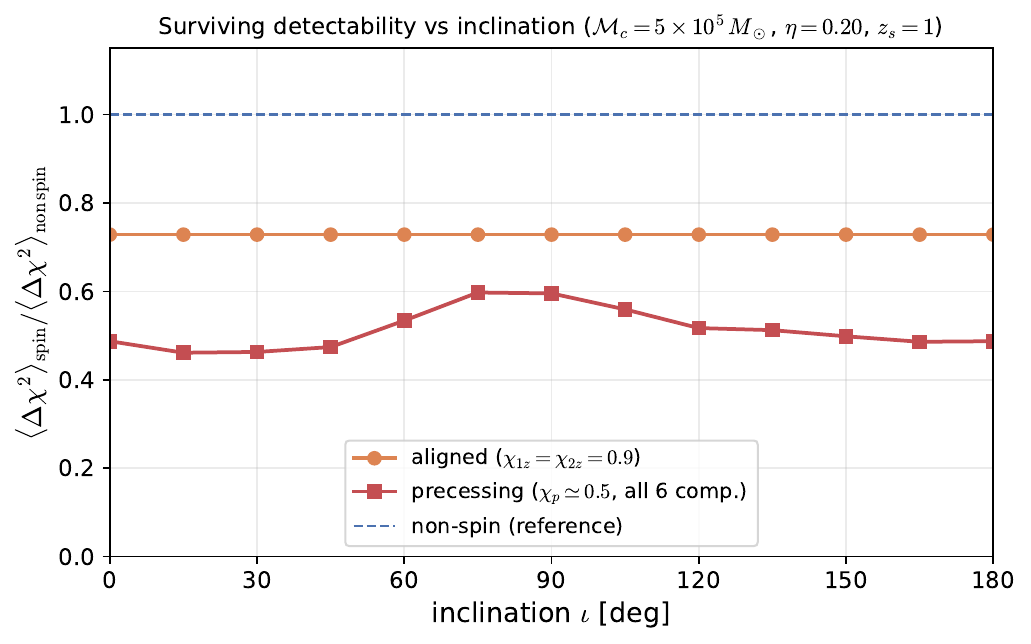}
    \caption{Surviving detectability $\langle\Delta\chi^2\rangle$ as a function of inclination $\iota$, normalised to the non-spin value at each $\iota$, for the high-spin aligned and fully precessing configurations. Aligned spins are orientation independent; precession produces an $\iota$-dependent reduction between $\simeq0.46$ and $\simeq0.60$.}
    \label{fig:spin_iota}
\end{figure}

\begin{figure}[t]
    \centering
    \includegraphics[width=0.60\textwidth]{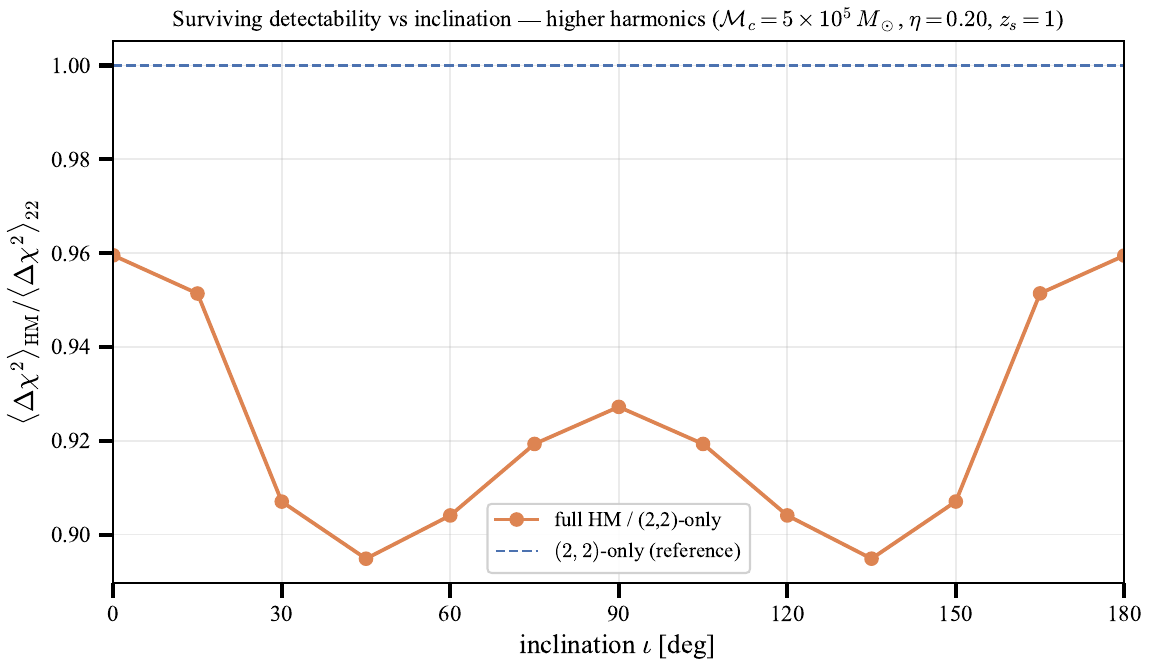}
    \caption{Surviving detectability $\langle\Delta\chi^2\rangle$ of the full-harmonic manifold relative to the $(2,2)$-only one as a function of inclination (fixed eigenmodes, quadrupole band). The $m\neq2$ harmonics vanish face-on, so the ratio approaches unity there; the largest reduction, $\simeq0.89$, occurs at intermediate inclinations $\iota\simeq45^\circ\,(135^\circ)$, with a partial recovery edge-on. The effect stays within $1$--$11\%$ at all orientations.}
    \label{fig:xphm_iota}
\end{figure}

The forecasts of the main text marginalise the Dark Timbre over the five-parameter binary manifold $\vect{\theta}=(\Mc,\eta,t_c,\phi_c,\log A)$ of an analytic, quadrupole-only IMR template~\cite{Ajith:2007kx} with sky-averaged response. The amplitude $\log A$ is fully degenerate with the luminosity distance and the sky-averaged antenna pattern, so marginalising over it also marginalises over $d_L$, sky position and polarisation; the inclination, the one extrinsic parameter that is not a pure amplitude rescaling once precession or higher harmonics are included, is fixed at $\iota=\pi/3$ and scanned explicitly below (Figs.~\ref{fig:spin_iota} and~\ref{fig:xphm_iota}). Real binaries, however, carry spins, enlarging the manifold to seven (aligned) or eleven (fully precessing) parameters, and radiate higher harmonics. This section quantifies both effects at a benchmark event ($\Mc=5\times10^{5}\,\Msun$, $\eta=0.20$, $z_s=1$, $5$\,yr) and validates the analytic template against them. We use the analytic template in the population forecasts because it is fast: the $(\Mc,z_s)$-grid and merger-rate scans of the main text require waveform derivatives and a covariance eigen-decomposition for every event, which is slower with the finite-difference waveform-model pipelines employed here.

\begin{figure}[t]
    \centering
    \includegraphics[width=0.60\textwidth]{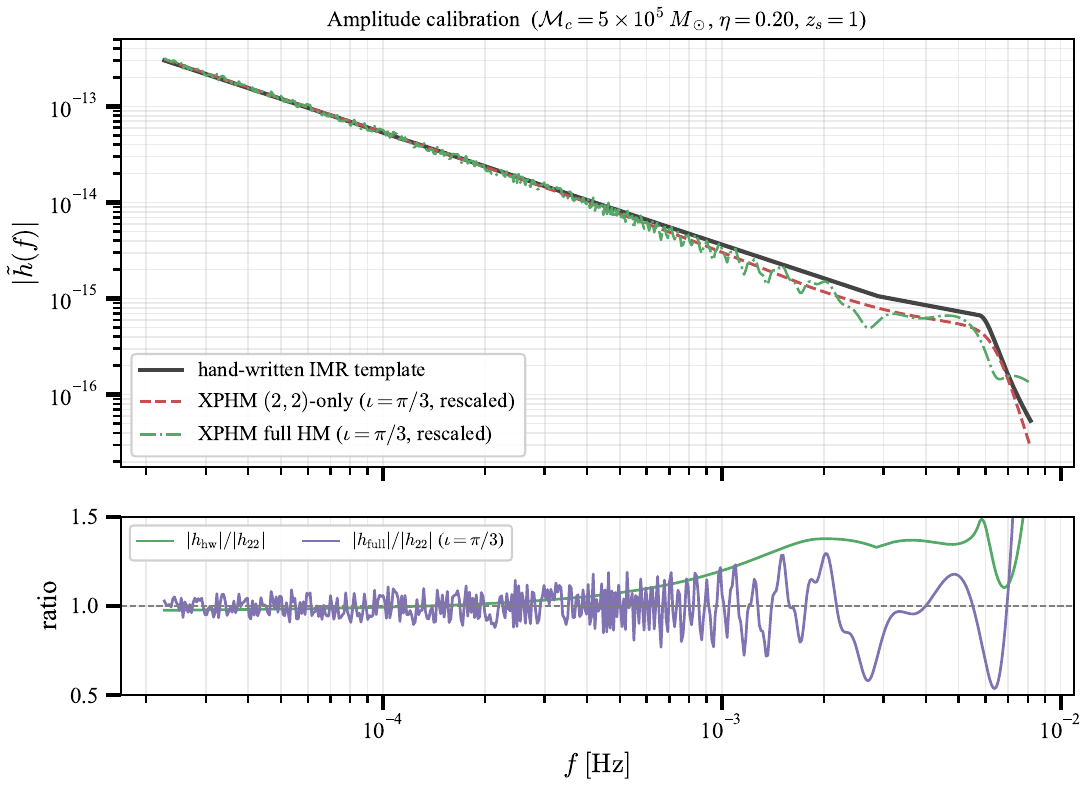}
    \caption{Amplitude calibration at the benchmark event. Top: the analytic IMR template and the calibrated \textsc{IMRPhenomXPHM} waveform in its $(2,2)$-only and full-harmonic variants, both shown at the benchmark inclination $\iota=\pi/3$ and rescaled by a common constant so that the $(2,2)$-only curve overlays the template in the inspiral --- the splitting between the two \textsc{XPHM} curves is then pure harmonic content. The calibration itself is performed face-on, where the inclination is a pure amplitude factor for the quadrupole, and matches the template to $5\%$ (RMS) over the inspiral band. Bottom: the calibration ratio $|h_{\rm hw}|/|h_{22}|$ and the harmonic ratio $|h_{\rm full}|/|h_{22}|$, whose mode-beating oscillations grow towards the merger--ringdown.}
    \label{fig:xphm_calibration}
\end{figure}

\emph{Template accuracy ---} We first consider the \textsc{IMRPhenomPv2}~\cite{Hannam:2013oca,Schmidt:2014iyl,Husa:2015iqa,Khan:2015jqa} waveform model (through \textsc{ripple}~\cite{Edwards:2023sak}), which includes spin precession but no higher-order modes, amplitude-calibrated to the analytic template ($5\%$ RMS over the inspiral band). At fixed Dark Timbre eigenmodes $(\sigma_n,\psi_n)$ (Sec.~\ref{sec:detstat}), its non-spinning limit reproduces the baseline catalogue size $N_{\rm bin}$ to $1\%$ (Table~\ref{tab:spin}); the $(2,2)$-only variant of \textsc{IMRPhenomXPHM}~\cite{Pratten:2020ceb} (through \textsc{lalsimulation}), calibrated in the same way (Fig.~\ref{fig:xphm_calibration}), reproduces it to $3\%$. The analytic template is therefore accurate enough for the forecasts: every difference reported below comes from the added physics --- spins or harmonics --- not from the template itself.

\begin{figure}[t]
    \centering
    \includegraphics[width=0.82\textwidth]{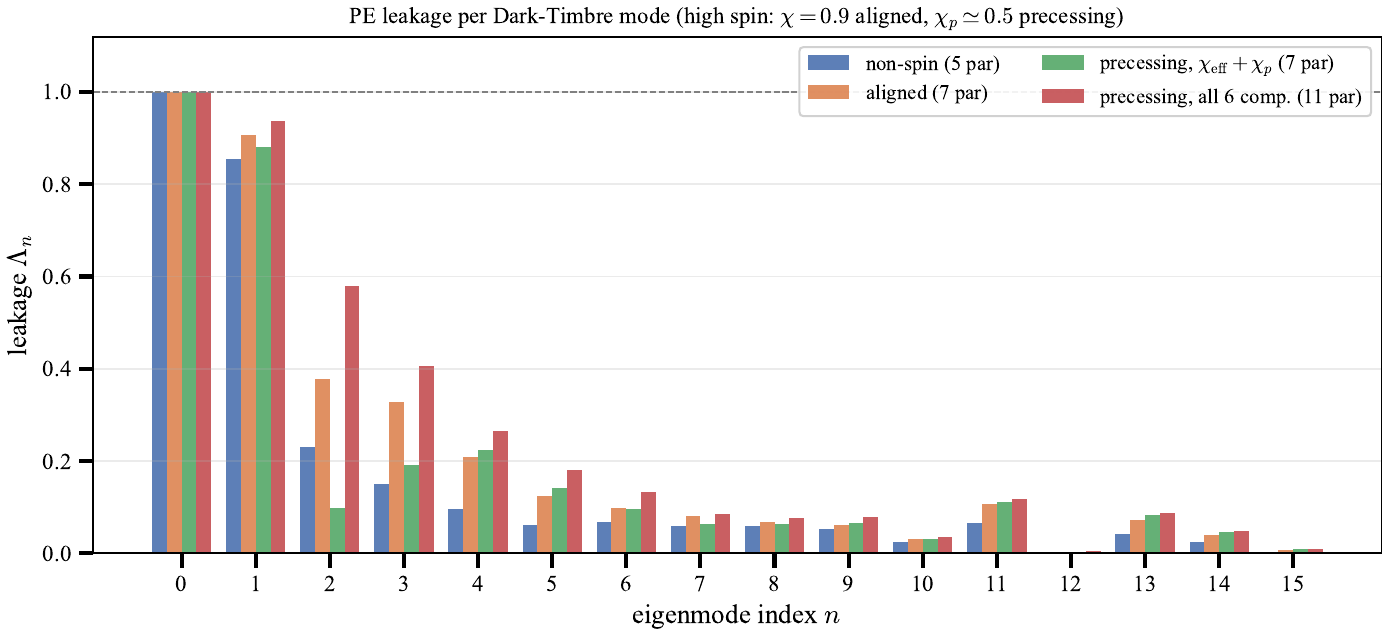}
    \caption{Leakage $\Lambda_n$ of each Dark Timbre eigenmode into the parameter manifold, for the four manifolds at the high-spin configuration. The geometric ($n=0$) mode is fully degenerate with the binary parameters in every case; the per-event significance is dominated by the next mode ($n=1$, arrow), whose leakage grows from $\Lambda_1\simeq0.85$ (non-spin) towards unity as spin parameters are added. Aligned spins and the effective $(\chi_{\rm eff},\chi_p)$ manifold leave most modes nearly unchanged; the conservative eleven-parameter case absorbs a large share of the low-order surviving power.}
    \label{fig:spin_ladder}
\end{figure}

\emph{Spins and precession ---} Enlarging the parameter manifold increases the leakage $\Lambda_n$ of the Dark Timbre eigenmodes, and at this event the significance is carried almost entirely by a single surviving mode [$N_{\rm eff}\equiv(\sum_n s_n)^2/\sum_n s_n^2\simeq1$; Fig.~\ref{fig:spin_ladder}], so the loss depends on how directly the new spin directions overlap it. Holding the eigenmodes fixed at the non-spinning template (the conservative choice), aligned spins reduce the per-event significance by $18$--$31\%$ ($N_{\rm bin}\times1.5$--$2.1$) and the fully precessing eleven-parameter marginalisation by $45$--$52\%$ ($N_{\rm bin}\times3.3$--$4.3$), for moderate to high spins (Table~\ref{tab:spin}). Marginalising only over the two effective combinations $(\chi_{\rm eff},\chi_p)$ that parameter estimation actually constrains gives $-6\%$ to $-27\%$; we quote it for reference, the physical cases being the aligned and fully precessing ones. The dependence on inclination is shown in Fig.~\ref{fig:spin_iota}: for aligned spins the surviving power is orientation independent ($\simeq0.7$ of the non-spin value), while for the fully precessing case it varies between $\simeq0.46$ (face-on) and $\simeq0.60$ (edge-on).

\begin{table}[h!]
\centering
\caption{Spin-precession ladder at the benchmark event ($\Mc=5\times10^{5}\,\Msun$, $\eta=0.20$, $z_s=1$, $5\,$yr; fixed eigenmodes, $\iota=\pi/3$). For each manifold we list the surviving fraction $f_{\rm surv}=\sum_n s_n/\sum_n\sigma_n^2$, the surviving power $\langle\Delta\chi^2\rangle=\sum_n s_n$, the per-event significance $\rho$, and the catalogue size $N_{\rm bin}=(S/\rho)^2$ for $S=2,5$. The non-spinning \textsc{IMRPhenomPv2} row matches the analytic-template baseline to $1\%$. The $(\chi_{\rm eff},\chi_p)$ rows marginalise only over the two effective spin combinations constrained by parameter estimation and are shown for reference; the physical cases are the aligned (7-parameter) and fully precessing (11-parameter) ones.}
\label{tab:spin}
\begin{tabular}{|l|c|c|c|c|c|c|}
\hline
Manifold & \#par & $f_{\rm surv}$ & $\langle\Delta\chi^2\rangle$ & $\rho$ & $N_{\rm bin}(2\sigma)$ & $N_{\rm bin}(5\sigma)$\\
\hline
(baseline) & 5 & $6.9\times10^{-4}$ & $0.967$ & $0.283$ & $50$ & $312$\\
\textsc{IMRPhenomPv2} non-spin & 5 & $6.9\times10^{-4}$ & $0.971$ & $0.281$ & $51$ & $316$\\
\hline
\multicolumn{7}{|l|}{\textit{moderate spin} ($\chi_{\rm eff}=0.31$, $\chi_p=0.35$; aligned $\chi=0.5$)}\\
\hline
\quad aligned & 7 & $5.7\times10^{-4}$ & $0.809$ & $0.231$ & $75$ & $469$\\
\quad fully precessing & 11 & $4.2\times10^{-4}$ & $0.586$ & $0.155$ & $167$ & $1044$\\
\quad $(\chi_{\rm eff},\chi_p)$ reference & 7 & $5.3\times10^{-4}$ & $0.744$ & $0.204$ & $96$ & $601$\\
\hline
\multicolumn{7}{|l|}{\textit{high spin} ($\chi_{\rm eff}=0.43$, $\chi_p=0.50$; aligned $\chi=0.9$)}\\
\hline
\quad aligned & 7 & $5.0\times10^{-4}$ & $0.704$ & $0.194$ & $107$ & $667$\\
\quad fully precessing & 11 & $3.7\times10^{-4}$ & $0.523$ & $0.136$ & $216$ & $1349$\\
\quad $(\chi_{\rm eff},\chi_p)$ reference & 7 & $6.5\times10^{-4}$ & $0.916$ & $0.264$ & $57$ & $359$\\
\hline
\end{tabular}
\end{table}

\emph{Why spins do not always degrade the signal ---} The fixed-eigenmode ladder is conservative because the Dark Timbre basis is not optimised for the spinning waveform. Spins change the merger amplitude and hence the weight $W(f)=|h|^2/S_n$ that defines the eigenmodes; rebuilding $\sigma_n,\psi_n$ self-consistently from the calibrated \textsc{IMRPhenomPv2} amplitude, the reshaped weight overlaps the lensing power more favourably and compensates the manifold cost: at the benchmark event the fully precessing eleven-parameter catalogue size becomes $0.84\times$ the self-consistent non-spin value --- slightly \emph{better} than no spin --- and the $(\chi_{\rm eff},\chi_p)$ case $0.36\times$. Two opposite effects compete: the enlarged manifold always absorbs surviving power, while the spin-modified merger amplitude adds raw lensing power, and the net outcome ranges from a factor-few penalty (fixed basis, flat priors on all six spin components) to a mild enhancement (self-consistent basis). We performed the self-consistent rerun at the benchmark event only: it requires re-solving the eigenproblem with waveform-model derivatives for every configuration, which is too expensive for the catalogue scans --- the same reason the forecasts use the analytic template. The main-text forecasts are therefore mildly conservative with respect to spin.

Note that \textsc{IMRPhenomPv2} only includes the quadrupolar emission $\ell=2$, thus neglecting additional information and making these results conservative; we quantify the harmonic content next.

\begin{figure}[t]
    \centering
    \includegraphics[width=0.82\textwidth]{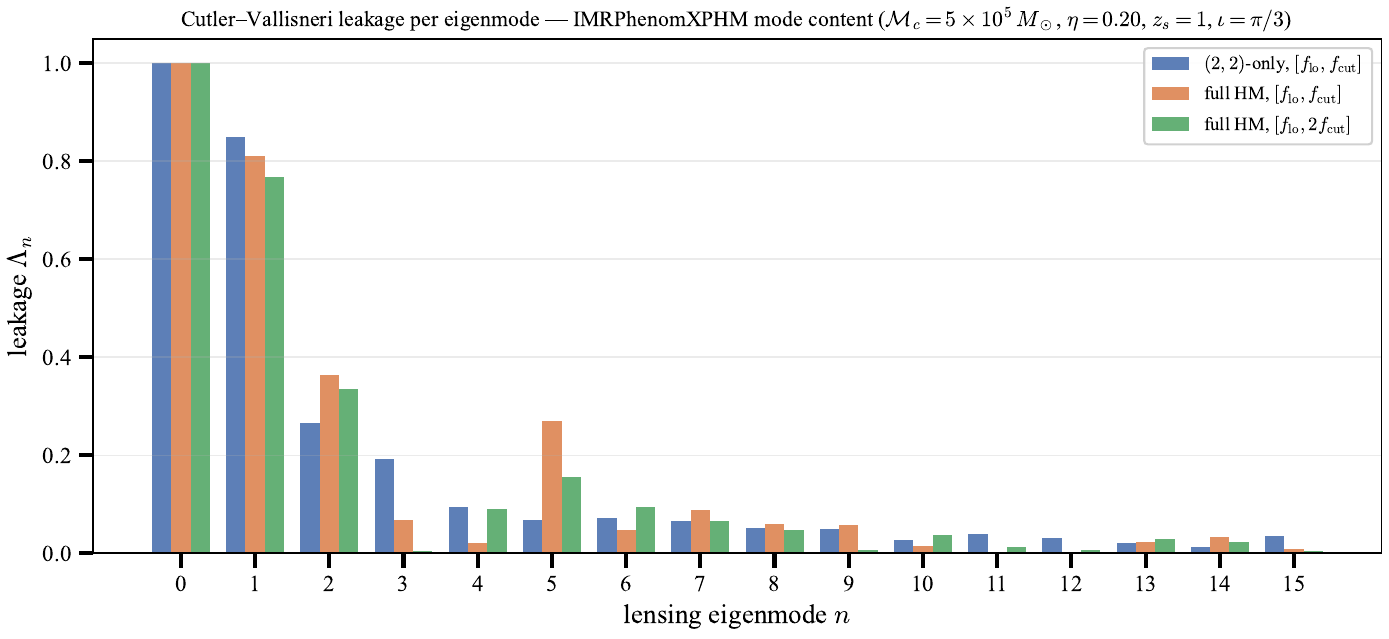}
    \caption{Leakage $\Lambda_n$ of each Dark Timbre eigenmode into the five-parameter manifold for the three configurations of Table~\ref{tab:harmonics}. Each configuration is decomposed in its own self-consistent eigenbasis, so the index $n$ labels a \emph{different} eigenfunction in each case and $\Lambda_n$ at fixed $n$ need not be monotonic between bands (e.g.\ $n=4,6$); the basis-independent statement is the significance $\rho^2=\tfrac14\sum_n\sigma_n^4(1-\Lambda_n)^2$, which sets the row ratios of Table~\ref{tab:harmonics}. On the extended band the low-order modes leak less and more modes survive, driven by the $(3,3)$ merger--ringdown peak.}
    \label{fig:xphm_ladder}
\end{figure}

\emph{Higher harmonics ---} We repeat the analysis with \textsc{IMRPhenomXPHM}~\cite{Pratten:2020ceb}, whose harmonic content $(\ell,|m|)=(2,2),(2,1),(3,3),(3,2),(4,4)$ can be restricted mode by mode, so the $(2,2)$-only and full-harmonic variants of the \emph{same} model are compared directly. At fixed frequency band the mode content is a $\lesssim10\%$ effect: with fixed eigenmodes the full content lowers the per-event significance by $10\%$ ($N_{\rm bin}\times1.24$); with self-consistent eigenmodes it instead \emph{raises} it by $5\%$ ($N_{\rm bin}\times0.91$; Table~\ref{tab:harmonics}); and across all inclinations the surviving power stays within $1$--$11\%$ of the $(2,2)$-only value (Fig.~\ref{fig:xphm_iota}). The corresponding leakage patterns are shown in Fig.~\ref{fig:xphm_ladder}.

\begin{table}[t]
\centering
\caption{Mode-content ladder at the benchmark event ($\Mc=5\times10^{5}\,\Msun$, $\eta=0.20$, $z_s=1$, $5\,$yr; five-parameter manifold, $\iota=\pi/3$), in the self-consistent convention where each configuration rebuilds the eigenmodes from its own $W(f)=|h|^{2}/S_n$. Because the $\iota=\pi/3$ amplitude suppresses the weight, the absolute $N_{\rm bin}$ are not comparable with Table~\ref{tab:spin} (fixed eigenmodes); the meaningful quantities are the ratios between rows. The $(2,2)$-only variant reproduces the analytic-template baseline to $3\%$ at fixed eigenmodes.}
\label{tab:harmonics}
\begin{tabular}{|l|c|c|c|c|c|c|}
\hline
Configuration & Band & $f_{\rm surv}$ & $\langle\Delta\chi^2\rangle$ & $\rho$ & $N_{\rm bin}(2\sigma)$ & $N_{\rm bin}(5\sigma)$\\
\hline
\textsc{XPHM} $(2,2)$-only & $[f_{\rm lo},f_{\rm cut}]$ & $7.5\times10^{-4}$ & $0.224$ & $0.064$ & $972$ & $6077$\\
\textsc{XPHM} full harmonics & $[f_{\rm lo},f_{\rm cut}]$ & $8.0\times10^{-4}$ & $0.234$ & $0.067$ & $887$ & $5542$\\
\textsc{XPHM} full harmonics & $[f_{\rm lo},2f_{\rm cut}]$ & $9.9\times10^{-4}$ & $0.295$ & $0.088$ & $522$ & $3262$\\
\hline
\end{tabular}
\end{table}

\begin{figure}[t]
    \centering
    \includegraphics[width=0.6\textwidth]{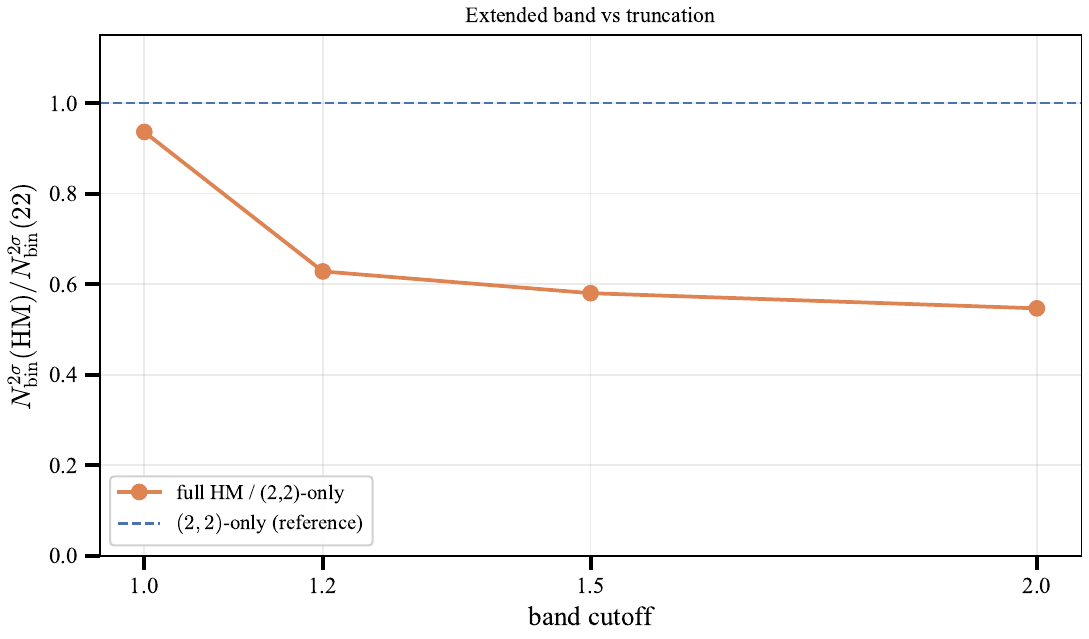}
    \caption{Catalogue size of the full-harmonic analysis relative to the $(2,2)$-only one, as the band is truncated at $X f_{\rm cut}$ (self-consistent $W(f)$, five-parameter manifold). Most of the extended-band gain arrives by $X=1.2$, driven by the $(3,3)$ merger--ringdown peak.}
    \label{fig:xphm_trunc}
\end{figure}

The dominant effect of the harmonics is not the mode content but the accessible frequency band. The $(\ell,m)$ harmonic radiates up to $\sim(m/2)$ times the quadrupole cut-off, so only the harmonics populate $f>f_{\rm cut}$ --- precisely where the lensing power keeps growing. Extending the band to $2f_{\rm cut}$, the harmonics above $f_{\rm cut}$ carry only $1.9\%$ of the total $\mathrm{SNR}^{2}$ (still $\mathrm{SNR}\simeq800$ at this loud event), yet the catalogue size drops to $0.54\times$ the $(2,2)$-only value ($+36\%$ in $\rho$; third row of Table~\ref{tab:harmonics}). A band-truncation ladder (Fig.~\ref{fig:xphm_trunc}) localises the gain: $\sim80\%$ of it accumulates already in $[f_{\rm cut},1.2\,f_{\rm cut}]$ --- the merger--ringdown peak of the $(3,3)$ mode, a calibrated region of the waveform model --- rather than in the far tail near $2f_{\rm cut}$ where the model is least reliable.

In summary: (i) the analytic template of the main text is validated to $1$--$3\%$ by the non-spinning and $(2,2)$-only limits of the waveform models; (ii) spins make the Dark Timbre somewhat harder to detect at fixed eigenmodes (up to $N_{\rm bin}\times4.3$ in the most conservative eleven-parameter case), but the spin-modified merger amplitude compensates the manifold cost once the eigenmodes are rebuilt self-consistently, leaving the detectability comparable to --- and occasionally better than --- the non-spinning forecasts; (iii) at fixed band the harmonic content changes the catalogue size by $\sim10\%$, and the larger gain comes from the band extension above the quadrupole cut-off, $N_{\rm bin}\times0.54$, which we quote as an optimistic upper bound: it is evaluated for a single event at $\iota=\pi/3$ (face-on binaries, favoured by detection selection, gain nothing) and the validation of the linear approximation in Sec.~\ref{sec:validity} does not yet cover $f>f_{\rm cut}$.

\newpage
\section{Validity of the linear approximation}\label{sec:validity}

The Dark Timbre signal is the decomposition of the stochastic diffractive fluctuation $\Delta F(w)$. The total signal is computed in the linear (Born) regime from the single-halo response $\Delta F_1$ Eq.~\eqref{eq:DF1} of the halo population Eq.~\eqref{eq:DFsplit}, and its detectability is computed from the \emph{analytic Gaussian covariance} of Eq.~\eqref{eq:cov_def}. Two assumptions underlie this: that the halo-by-halo \emph{linear} superposition reproduces the full, non-perturbative wave-optics amplification; and that this analytic, infinite-volume covariance, rather than explicit Poisson realisations of the halos on a finite box, is the correct description of the detectable signal.

We test the first directly against the development version of the \glow{} code~\cite{Villarrubia-Rojo:2024xcj}, which evaluates the amplification factor $F(f)$ using full-wave optics via 2D contour integration method. This allows us to validate the linearization of the signal. As shown in Section~\ref{sec:multiplane}, multi-plane lensing can be reduced to an effective single-plane description through the effective lensing potential~\eqref{eq:Psieff}. By plugging this potential into \glow, which is designed for single-plane lensing, we can obtain the wave-optics amplification factor beyond the perturbative, linearised analysis. Although the nonlinear structure of multi-plane lensing is more involved than that of single-plane lensing, the two are similar in the sense that their nonlinear corrections can be organised perturbatively as powers of $\Psi$. Therefore, validating the linear approximation in the single-plane case provides strong evidence that nonlinear effects remain small also in the multi-plane case. The tests use ten fixed halo configurations sampled at source redshifts $z_s\in[0.26,0.46]$, with halo masses $M\in[10,10^{8}]\,M_\odot$ and reference frequency $f_{\min}=5\times10^{-4}\,$Hz. 
\begin{figure*}[h!]
    \centering
    \includegraphics[width=0.8\textwidth]{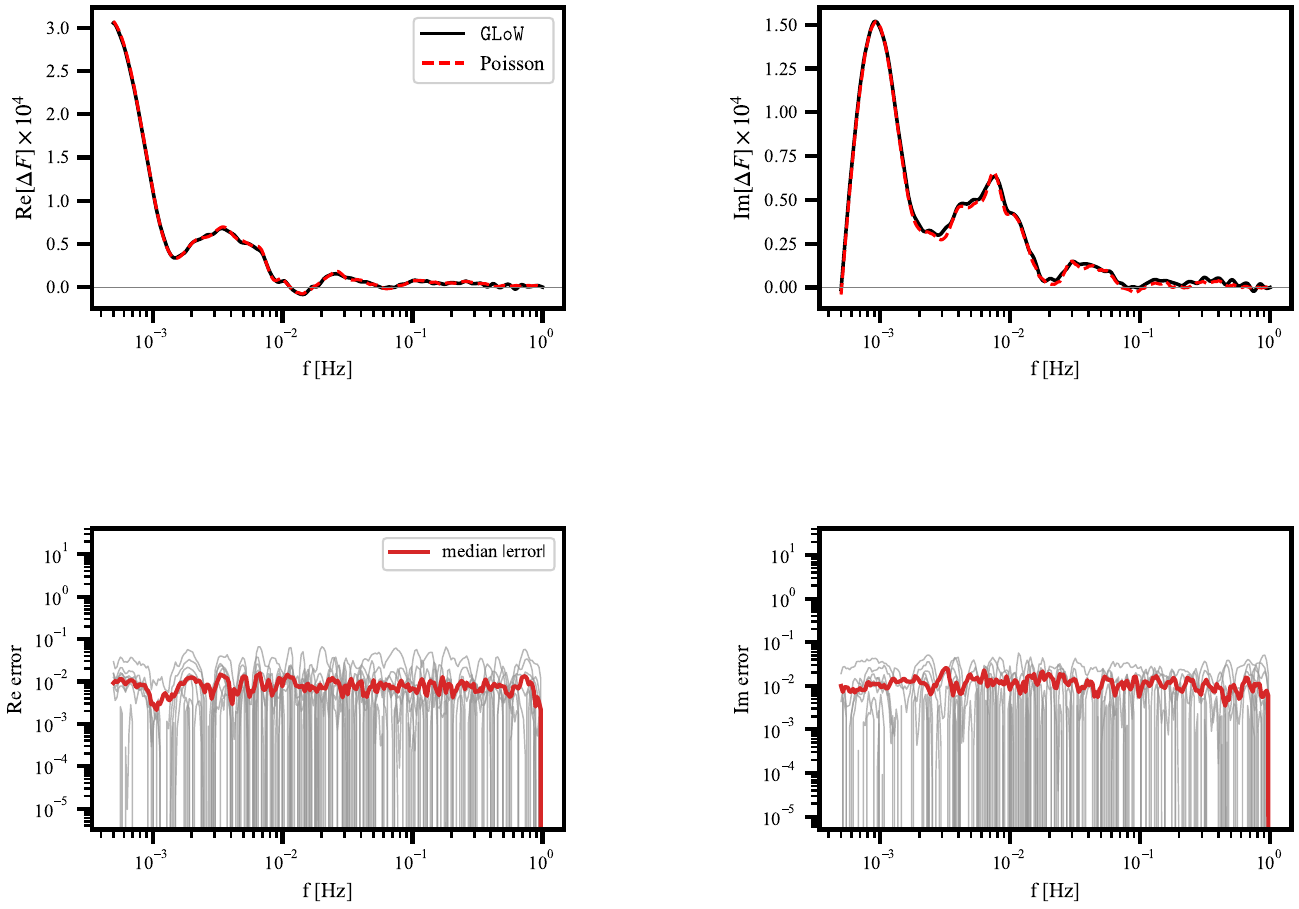}
    \caption{Linear (Born) Poisson method versus the full wave-optics code \glow\ on ten identical halo catalogues. \emph{Top:} real and imaginary parts of the diffractive amplification $\Delta F(f)$ for a representative configuration ($N_{\rm halos}=477$, $z_s=0.45$, $z_l=0.2$); \glow\ (solid) and the linear method (dashed) are aligned to their geometric-optics value at $f_{\max}$. \emph{Bottom:} relative error $(\Delta F_{\rm lin}-\Delta F_{\glow})/\max_f|\Delta F_{\glow}|$ for each of the ten runs (grey) and the median absolute error (red).}
    \label{fig:glow_poisson}
\end{figure*}

Figure~\ref{fig:glow_poisson} overlays the real and imaginary parts of $\Delta F(f)$ from the two methods on a representative configuration (top row), together with the relative error of all ten runs (bottom row). To make the comparison clean we treat the mean field exactly as \glow\ does: the linear $\Delta F$ carries \emph{no} per-halo disk-average subtraction, and the geometric/mean field is removed instead by aligning both methods to their value at $f_{\max}$ (\glow's $\sqrt{\mu}$ normalisation). With this matched convention, the linear method reproduces the full wave-optics result oscillation by oscillation across the \emph{entire} band: the median error is $\sim1$--$2\%$ at all frequencies, low $f$ included, with individual runs staying within a few percent. We stress that this holds down to the lowest frequencies: had we instead applied the disk-average subtraction per halo (the superseded normalisation), the same comparison would show a spurious growth of the error to tens of percent below $w\simeq4$, but that reflects only the mismatch between the two ways of enforcing $\langle F\rangle=1$ --- the per-halo disk average versus \glow's $\sqrt{\mu}$ --- and \emph{not} a breakdown of the linear approximation, as removing it here confirms. The linear, Born treatment is thus an essentially exact description of the wave-optics amplification for these weak lenses across the whole diffractive band.

\begin{figure}[t]
    \centering
    \includegraphics[width=0.6\columnwidth]{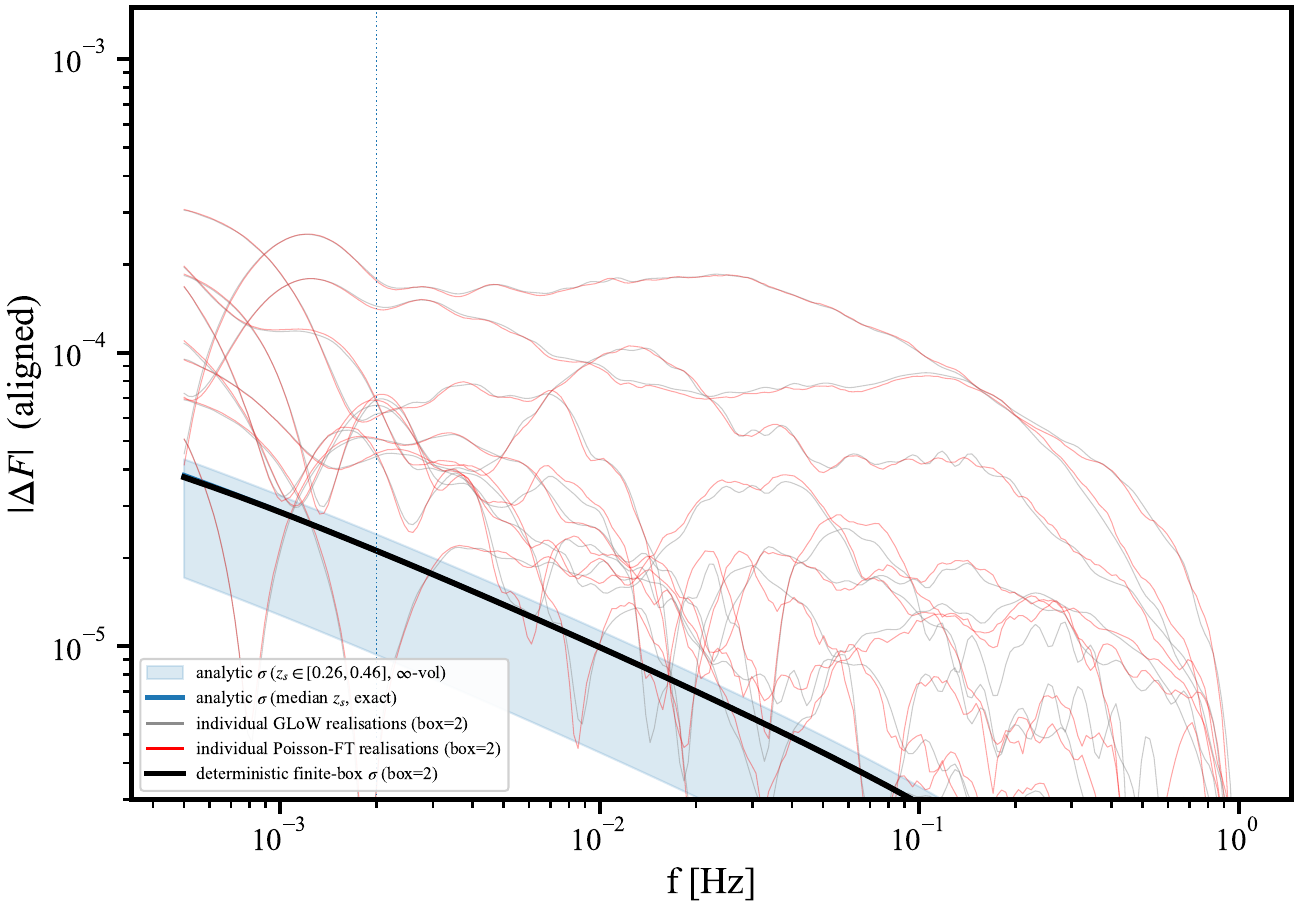}
    \caption{The finite box is adequate, but explicit realisations are an unreliable estimator of the variance. Thin lines: $|\Delta F(f)|$ for the ten \glow\ (grey) and ten linear (red) realisations on their fixed two-Fresnel-scale catalogues; they overlap pairwise (linear $=$ \glow) yet scatter over more than a decade, because rare massive halos dominate the variance. Thick black: the \emph{deterministic} finite-box variance, which coincides with the exact infinite-volume $\sigma_{\Delta F}$ of Eq.~\eqref{eq:cov_def} (blue line, with the $z_s$-range band). The analytic covariance is exact and smooth; the explicit Monte Carlo estimate is heavy-tailed and noisy, which is why the detectable signal is computed from the former.}
    \label{fig:finite_volume_lowf}
\end{figure}

A distinct question is whether the \emph{finite sampling box} biases the variance of the signal; it does not. The analytic covariance of Eq.~\eqref{eq:cov_def} is the \emph{exact} infinite-volume variance of the diffractive signal. The same integral truncated at a finite box radius $R$ is the variance an explicit catalogue of that size would have; because the geometry-aligned single-halo response $X(r_c)\equiv\Delta F_1(w;r_c)-\Delta F_1(w_{\max};r_c)$ decays within $r_c\lesssim2$ (its long-range geometric tail cancels in the alignment), this deterministic finite-box variance already equals the exact infinite-volume $\sigma_{\Delta F}$ for $R=2$ to better than $5\%$, and is independent of $R$ thereafter: \emph{there is no finite-volume deficit}. What is true is that estimating the variance from explicit Poisson realisations is unreliable. The variance over the mass range $[10,10^{8}]\,M_\odot$ is dominated by the rare, heavy halos, so the $|\Delta F|$ of a feasible number of catalogues scatters over more than a decade around the smooth analytic value (Fig.~\ref{fig:finite_volume_lowf}). Also each per-halo $J_0(k\,r_c)$ kernel oscillates with a period $\sim\pi/r_c$ that shrinks with impact parameter, so a coarse wavenumber grid under-resolves it and spuriously inflates the estimate. \glow\ and the linear method coincide realisation by realisation, confirming once more that the linear treatment reproduces the full wave optics, but their root-mean-square over ten catalogues is a heavy-tail-biased, noisy estimate, not the variance.

These are the reasons the Dark Timbre detectability of the main text is built \emph{entirely} from the analytic Gaussian covariance Eq.~\eqref{eq:cov_def} and never from explicit Poisson halos. The detectable signal is carried by the many low-mass halos that populate each Fresnel volume in large numbers; by the central limit theorem their summed $\Delta F$ is Gaussian and \emph{fully} specified by the covariance Eq.~\eqref{eq:cov_def}. The explicit Poisson piece of Eq.~\eqref{eq:DFsplit} only becomes relevant for the rare, heavy halos, whose contribution is the non-Gaussian, heavy-tailed \emph{geometric} mode --- real, nearly frequency independent, and degenerate with the binary parameters, so that it is projected out as the $n=0$ mode of the Karhunen--Lo\`eve expansion and does not enter the test statistic. The mass split of Eq.~\eqref{eq:DFsplit} therefore separates the signal cleanly: the numerous light halos, which carry the detectable diffractive power, are treated with the exact analytic covariance; the rare heavy halos, which carry only the parameter-degenerate geometric mode, are the \emph{only} ones sampled explicitly, and there the handful of lenses involved makes the box and quadrature issues immaterial.

The tests of this section establish the three results that underpin our detectability forecast. 
\begin{itemize}
    \item On identical halo configurations the linear (Born) method reproduces the full \glow\ wave-optics amplification to $\sim1$--$2\%$ across the \emph{entire} diffractive band, low $f$ included, once the mean field is treated as in \glow\ (no disk-average subtraction; alignment to $f_{\max}$). This is consistent with the production convention of Eq.~\eqref{eq:DF1mean}, whose deterministic mean-field term is constant up to $\mathcal{O}(2/wR^2)$ and drops out of the alignment; the low-$f$ degradation obtained with the superseded per-halo disk-average subtraction was a mean-field convention artefact, not a breakdown of the linear approximation.
    \item The detectable signal is therefore computed from the analytic Gaussian covariance Eq.~\eqref{eq:cov_def}, which evaluates the impact-parameter integral in closed form (exact, smooth, noiseless), rather than from explicit Poisson realisations, which are a heavy-tail-noisy and numerically delicate estimator of the same variance. 
    
    \item The split is physically clean: the diffractive power that is \emph{measurable} is carried by the many light halos --- Gaussian by the central limit theorem and fully specified by the covariance --- whereas the rare massive halos contribute only the geometric ($n=0$) mode, which is real, frequency-independent and degenerate with the binary parameters; this is exactly the magnification that conventional gravitational-wave strong-lensing studies isolate, and it drops out of our test statistic.
\end{itemize}

\end{document}